\documentstyle[aps,pre,floats,epsf,eqsecnum]{revtex}
\begin{document}
\advance\textheight by 0.2in \draft

\twocolumn[\hsize\textwidth\columnwidth\hsize\csname@twocolumnfalse%
\endcsname

\title{Novel non-equilibrium critical behavior in unidirectionally
       coupled stochastic processes}

\author{Yadin Y. Goldschmidt$^1$, Haye Hinrichsen$^2$, 
        Martin Howard$^{3,*}$, and Uwe C. T\"auber$^{4,*}$}

\address{$^1$ Department of Physics and Astronomy, 
         University of Pittsburgh, Pittsburgh PA 15260, U.S.A. \\ 
         $^2$ Max-Planck-Institut f\"ur Physik komplexer Systeme,
         N\"othnitzer Stra\ss e 38, D-01187 Dresden, Germany \\ 
         $^3$ CATS, The Niels Bohr Institute, Blegdamsvej 17, 
         2100 Copenhagen \O, Denmark \\ 
         $^4$ Institut f\"ur Theoretische Physik, 
         Technische Universit\"at M\"unchen,
         James-Franck-Stra\ss e, D-85747 Garching, Germany \\}

\date{\today} 

\maketitle

\begin{abstract}
  Phase transitions from an active into an absorbing, inactive state
  are generically described by the critical exponents of directed
  percolation (DP), with upper critical dimension $d_c = 4$.  In the
  framework of single-species reaction-diffusion systems, this
  universality class is realized by the combined processes $A \to A +
  A$, $A + A \to A$, and $A \to \emptyset$.  We study a hierarchy of
  such DP processes for particle species $A, B,\ldots$, {\em
  unidirectionally} coupled via the reactions $A \to B, \ldots$ (with
  rates $\mu_{AB}, \ldots$).  When the DP critical points at all
  levels coincide, {\em multicritical} behavior emerges, with density
  exponents $\beta_i$ which are markedly reduced at each hierarchy
  level $i \geq 2$.  This scenario can be understood on the basis of
  the mean-field rate equations, which yield $\beta_i = 1/2^{i-1}$ at
  the multicritical point.  Using field-theoretic renormalization
  group techniques in $d = 4 - \epsilon$ dimensions, we identify a new
  crossover exponent $\phi$, and compute $\phi = 1 + O(\epsilon^2)$ in
  the multicritical regime (for small $\mu_{AB}$) of the second
  hierarchy level.  In the active phase, we calculate the fluctuation
  correction to the density exponent on the second hierarchy level,
  $\beta_2 = 1/2 - \epsilon/8 + O(\epsilon^2)$.  Outside the
  multicritial region, we discuss the crossover to ordinary DP
  behavior, with the density exponent $\beta_1 = 1 - \epsilon/6 +
  O(\epsilon^2)$.  Monte Carlo simulations are then employed to
  confirm the crossover scenario, and to determine the values for the
  new scaling exponents in dimensions $d \leq 3$, including the
  critical initial slip exponent.  Our theory is connected to specific
  classes of growth processes and to certain cellular automata, and
  the above ideas are also applied to unidirectionally coupled pair
  annihilation processes.  We also discuss some technical as well as
  conceptual problems of the loop expansion, and suggest some possible
  interpretations of these difficulties.

\end{abstract}

\pacs{PACS numbers: 64.60.Ak, 05.40.+j, 82.20.-w.}]


\section{Introduction}
\label{intro}


The notion of {\em universality} plays a central role in equilibrium
as well as in non-equilibrium statistical mechanics.  It was first
used by experimental physicists in order to describe the observation
that certain thermodynamic observables measured in different and
apparently unrelated equilibrium systems near a continuous phase
transition may exhibit the same type of singular
behavior~\cite{critph}.  It was, in fact, then realized that the
majority of equilibrium critical phenomena belong to very few {\em
universality classes} which are characterized by a certain set of
critical exponents.  In order to explain universality, various
theoretical approaches have been constructed, for example scale
invariance~\cite{scalin}, field-theoretic renormalization group
techniques~\cite{rgtech}, and the theory of conformal
invariance~\cite{coninv}, which predicts a series of universality
classes for two-dimensional critical systems.  Thus in equilibrium
statistical mechanics, especially in two dimensions, the concept of
universality seems to be well understood.

For systems far from equilibrium, however, the situation near dynamic
continuous phase transitions is less clear.  Non-equilibrium processes
are much harder to solve or even to characterize exactly since the
probability distribution cannot be obtained from an energy functional,
but has to be derived directly from the equations of motion.  In
addition, systems far from equilibrium are in general not conformally
invariant since there is no symmetry between spatial and temporal
degrees of freedom.  Nevertheless it appears that universality,
although probably in a weaker sense, may also play an important role
in non-equilibrium critical phenomena.  As in the case of equilibrium
physics the picture that emerges is that only a few distinct
universality classes seem to exist.  The known examples include phase
transitions in driven diffusive systems~\cite{dridif}, the power-law
decay in annihilation-coagulation
processes~\cite{coagul,annihi,masfth}, the ``parity-conserving''
dynamic transition for branching and annihilating random walks with
even offspring number~\cite{evbarw}, non-equilibrium roughening
transitions in growth models~\cite{gromod}, specifically in the KPZ
equation~\cite{kapezt}, and the critical points of directed
percolation~\cite{percol}, as described by Reggeon field
theory~\cite{regfth}, and of dynamic (isotropic) 
percolation~\cite{dynper}.  As a unifying theoretical framework is not
yet available, we are still far from a systematic classification of
non-equilibrium critical phenomena.  Therefore one important direction
of research is in fact to search for further unknown universality
classes.

Another direction, which is actually the objective of the present
paper, would be to investigate the known universality classes in more
complicated contexts.  The basic idea is to use several
non-equilibrium systems of a known universality class as building
blocks of a superior structure in which the systems are linked to each
other in a specific manner.  The question posed is whether these
systems can be combined in such a way that novel critical behavior
emerges.  In other words, is it possible to couple several
non-equilibrium systems of a given universality class in such a manner
that the resulting critical behavior is characterized by independent
critical exponents?

Such a model with {\em quadratically} coupled directed percolation
(DP) processes was recently investigated by Janssen, who found that
despite the apparent complexity of this coupled multi-species system,
the universality class of the active / absorbing transition was that
of DP itself \cite{mulcdp}.  In the present article we show that novel
critical behavior may, however, occur when several copies of the same
non-equilibrium process are {\em linearly} coupled in one direction
without feedback.  More precisely, we consider a linear hierarchy of
{\em unidirectionally} coupled copies $A$,$B$,$C,\ldots$ of the {\em
same} non-equilibrium system:
\begin{equation}
\label{CoupledProcesses}
A \rightarrow B \rightarrow C \rightarrow \ldots 
\end{equation}
The systems are coupled in such a way that the dynamical processes at
a certain level in the hierarchy depend on the state at the preceding
level but not vice versa.  For example, subsystem~$A$, the lowest
level in the hierarchy, is not influenced by the dynamics of~$B$
and~$C$ and thus it evolves independently as if the other hierarchy
levels did not exist.  Subsystem $B$ in turn is affected by~$A$ but
not by~$C$, and hence this is the first level in the hierarchy where
novel critical behavior might occur.  The hierarchy can be continued
to infinitely many levels.  However, because of the unidirectional
structure one can always truncate the hierarchy at some level without
affecting the temporal evolution at lower levels.  For example, we may
consider a two-level hierarchy $A \rightarrow B$ or a three-level
hierarchy $A \rightarrow B \rightarrow C$; in both cases the dynamics
of the subsystems $A$ and $B$ would be exactly the same.

The most interesting behavior of the composite system is expected when
the subsystems $A,B,C,\ldots$ themselves are close to criticality.
This usually happens when the isolated subsystems would undergo a
continuous non-equilibrium phase transition.  Let us assume that the
stochastic process under consideration is controlled by a single
parameter~$p$ with a phase transition taking place at $p = p_c$.  A
unidirectionally coupled hierarchy of such processes is thus
controlled by a sequence of independent control parameters
$p^{(A)},p^{(B)},p^{(C)},\ldots$, which means that the composite
system is described by a high-dimensional phase diagram.  When all
levels are critical (i.e. $p^{(A)} = p^{(B)} = p^{(C)} = \ldots =
p_c$), a complex interplay of long-range correlations is expected.  We
will refer to this special point in the phase diagram as a {\em
multicritical point}.  In the vicinity of this point the properties of
the entire system depend crucially on the direction from which it is
approached, resulting in interesting {\em multicritical behavior}.  In
particular we will consider the special case $p^{(A)} = p^{(B)} =
p^{(C)} = \ldots = p$, where the entire hierarchy is controlled by a
single parameter.

The outlined concept of unidirectionally coupled non-equilibrium
processes is quite general and may be applied to various dynamical
systems.  But, as we shall also see, this mechanism does not
necessarily lead to new universality classes in all cases, and we have
already mentioned the quadratically coupled DP processes \cite{mulcdp}
as one counterexample.  In the present work we will focus on
non-equilibrium processes which display a continuous phase transition
from a fluctuating into an absorbing state, i.e. a configuration which
once reached, cannot be escaped from.

The canonical example for a transition into an absorbing state is the
critical point of directed percolation~(DP).  In DP, sites of a
lattice are either occupied by a particle (active) or empty
(inactive).  The dynamic processes are that a particle can
self-destruct or produce an offspring at a neighboring empty site.  If
the rate for offspring production $p$ is very low, the system always
reaches a state without particles which is the absorbing state of the
system.  On the other hand, when $p$ exceeds a certain critical value
$p_c$, another steady state with a finite particle density exists on
the infinite lattice.  In between a continuous phase transition takes
place which is characterized by long-range power-law correlations.
Another example considered in the present work is the annihilation
process $A+A \rightarrow 0$~\cite{annihi,masfth} in which the particle
density decays as $t^{-d/2}$ in dimensions $d < 2$.  Here the
absorbing state (the empty lattice) is approached without the tuning
of any parameter, i.e. the process is ``critical'' by itself.

To construct a unidirectionally coupled hierarchy of such processes we
implement additional dynamical rules which allow each particle at a
given level in the hierarchy to induce the creation of a new particle
at the same lattice site of the next level.  This ensures that the
composite system still has an absorbing state, namely the empty state
without particles.  More precisely, a whole hierarchy of absorbing
subspaces is generated.  For example, if subsystem $A$ enters the
inactive state, it will never become active again and therefore the
dynamical processes are restricted to an absorbing subspace where only
$B,C,\ldots$ may fluctuate.  This subspace in turn contains another
absorbing subspace in which $B$ is inactive, etc.  As we will
demonstrate, such a hierarchy of systems with absorbing states coupled
by induced particle creation is characterized by a subset of novel
critical exponents.  It should be emphasized that the emergence of
novel critical behavior is related to the fact that the processes are
coupled in only one direction.  Even a very small feedback (e.g. $B
\rightarrow A, C \rightarrow B$) or cyclic closure (e.g. $A
\rightarrow B \rightarrow C \rightarrow A$) would destroy this new
feature.

In this article we present a detailed analysis of unidirectionally
coupled DP~\cite{ourprl}.  For the case of equal control parameters it
is observed that the asymptotic particle densities near the
multicritical point are characterized by different critical exponents
$\beta_A, \beta_B, \beta_C, \ldots$.  Since level $A$ evolves
independently, $\beta_A$ is just the usual density exponent of DP.  At
higher levels numerical estimates show that the density exponents are
considerably reduced compared to their DP values.  In Sec.~II we
discuss the mean-field theory of coupled DP which already explains why
the density exponents at higher levels are reduced.  It also allows us
to study crossover phenomena close to the multicritical point.
Mean-field theory is expected to hold above the DP critical spatial
dimension $d_c=4$.  For $d < d_c$ the numerically observed values for
the density exponents are much smaller, which means that the
mean-field results are strongly modified by fluctuation effects.  In
order to understand these reduced values, we recently derived the
critical exponents to one-loop order~\cite{ourprl} by means of a
field-theoretical renormalization group analysis, based on the
``Hamiltonian'' representation of the classical master
equation~\cite{masfth,jcardy,redfth}.  In Sec.~III we present these
calculations in detail, including a discussion of the diagonalized and
multicritical theories, respectively, the active phase, logarithmic
corrections at $d_c = 4$, crossover studies, and the critical behavior
at higher levels in the hierarchy.  The field-theoretical results are
supported by extensive numerical simulations in Sec.~IV, while in
Sec.~V various applications of coupled DP are demonstrated. The
examples of coupled annihilation and other closely related topics are
the subjects of Sec.~VI.  Finally, a critical discussion of some
technical as well as conceptual problems arising in the
field-theoretic approach are the subject of our conclusions in
Sec.~VII.

\section{Coupled DP processes: Mean-field approximation}
\label{meanf}


In the bulk of this paper, we shall study unidirectionally coupled
directed-percolation processes.  It is convenient to represent these
processes in the framework of reaction-diffusion systems.  The
starting point of the hierarchy of coupled systems is therefore the
following reaction scheme, which can be viewed as the prototype for
the active / absorbing state transitions in the directed-percolation
universality class~\cite{regfth}:
\begin{eqnarray}
\label{branch}
&A \to A + A \quad &{\rm with \ rate \ } \sigma_A \ , \\
\label{decay}
&A \to \emptyset \quad &{\rm with \ rate \ } \mu_A \ , \\
\label{coagul}
&A + A \to A \quad &{\rm with \ rate \ } \lambda_A \ .
\end{eqnarray}
We can immediately write down the corresponding rate equations for the
particle density $n_A(t)$; this description neglects fluctuation and
correlation effects, and therefore corresponds to a mean-field
approximation.  The average particle number is increased via the
branching reaction (\ref{branch}), and reduced via both the decay
(\ref{decay}) and coagulation (\ref{coagul}).  However, while the
first two processes occur spontaneously, with rates $\sigma_A$ and
$\mu_A$, respectively, and therefore the change in particle number is
proportional to the particle density itself, the coagulation reaction
requires that two particles $A$ meet on the same lattice site (in a
discrete representation), and hence the total particle loss due to the
process (\ref{coagul}) is proportional to the density squared.  This
yields the balance equation
\begin{equation}
\label{mfrate}
{\partial n_A(t) \over \partial t} = 
(\sigma_A - \mu_A) n_A(t) - \lambda_A n_A(t)^2 \ .
\end{equation}
As we are considering local reactions only, we may generalize this
mean-field equation slightly by considering a coarse-grained local
particle density $n_A(x,t)$, and supplementing Eq.~(\ref{mfrate}) with
a diffusion term,
\begin{equation}
\label{mfeq}
{\partial n_A(x,t) \over \partial t} = D \left( \nabla^2 - r_A \right)
n_A(x,t) - \lambda_A n_A(x,t)^2 \ .
\end{equation}
Here we have introduced the diffusion constant $D$, and defined $r_A =
(\mu_A - \sigma_A) / D$.

Obviously, the mean-field dynamic phase transition occurs at the point
$r_A = 0$, where the balance of gain and loss due to the processes
linear in $n_A$ changes sign.  For $r_A > 0$, the only stationary
state of Eq.~(\ref{mfeq}) is $n_A = 0$, and for $t \to \infty$ the
particle density will simply decay to zero according to $n_A(t) \to
e^{-D r_A t}$, because once the particle density has become
sufficiently small, the coagulation contribution can be neglected.
Furthermore, $n_A = 0$ represents an absorbing phase, because once
there are no particles left in the system, none of the processes
(\ref{branch})--(\ref{coagul}) can happen any longer --- hence all
fluctuations cease, and the system cannot escape from this state.  For
$r_A < 0$, on the other hand, there is another stationary state with
non-zero particle density
\begin{equation}
\label{staden}
n_A = D |r_A| / \lambda_A \ ,
\end{equation}
which can be viewed as the order parameter of the active phase.  In
the active state, the asymptotic density is approached exponentially
again, with the characteristic rate $D |r_A|$.  Precisely at the
transition, only the term proportional to $n_A^2$ survives in
Eq.~(\ref{mfeq}), which therefore becomes identical to the mean-field
rate equation for diffusion-limited coagulation or annihilation
\cite{coagul}.  The solution to Eq.~(\ref{mfrate}) then becomes
\begin{equation}
\label{crdec}
n_A(t) \sim 1 / t \ ,
\end{equation}
i.e., the density decays according to a power law at the critical
point.

Upon identifying $r_A = p_c - p$, where $p_c$ denotes the percolation
threshold, we can translate the above mean-field results to the
notation of directed percolation.  Eq.~(\ref{mfeq}) implies that the
characteristic length scale $\xi_\perp = |r_A|^{-1/2}$ diverges in the
vicinity of the transition,
\begin{equation}
\label{mfnup}
\xi_\perp \sim |r_A|^{-\nu_\perp} \ , \quad \nu_\perp = 1/2 \ .
\end{equation}
At the critical point, the exponential decay rates vanish, and the
characteristic frequency becomes diffusive, $\omega_c \sim D q^2$;
hence
\begin{equation}
\label{mfz}
\omega_c \sim q^z \ , \quad z = 2 \ .
\end{equation}
Equivalently, upon approaching the critical point, the characteristic
time scale diverges according to
\begin{equation}
\label{mfnupar}
\xi_\parallel \sim |r_A|^{-\nu_\parallel} \ , \quad 
\nu_\parallel \equiv z \nu_\perp = 1 \ .
\end{equation}
Also, at $p = p_c$,
\begin{equation}
\label{mfalpha}
n_A(t) \sim t^{-\alpha} \ , \quad \alpha = 1 \ .
\end{equation}
Finally, in the active phase near the transition, the order parameter
grows as
\begin{equation}
\label{mfbeta}
n_A \sim |r_A|^{\beta} \ , \quad \beta = 1 \ .
\end{equation}
Notice that the exponents $\alpha$ and $\beta$ are related, provided
the following scaling relation holds,
\begin{equation}
\label{scaling}
n_A(r_A,x,t) = |r_A|^{\beta} {\hat n}_A(x/\xi_\perp, t/\xi_\parallel)
\ .
\end{equation}
For then at $p = p_c$, ${\hat n}_A(0,y) \sim y^{-\beta /
\nu_\parallel}$ in the limit $y \to \infty$, in order for the
$|r_A|$-dependence to cancel, and thus
\begin{equation}
\label{scarel}
\beta \equiv \nu_\parallel \alpha \equiv z \nu_\perp \alpha \ .
\end{equation}
The above relations therefore define {\em three} independent critical
exponents.  Alternatively, the third independent exponent may be
swapped for $\eta_\perp$, which characterizes how the equal-time pair
correlation function decays at the critical point $r_A = 0$, $G(|{\bf
x}|) \propto 1/|{\bf x}|^{d + z - 2 + \eta_\perp}$.

While the scaling relation (\ref{scarel}) remains valid also below the
upper critical dimension, which for DP turns out to be $d_c=4$ (see
Sec.~\ref{renorm}), the exponent values will become modified for $d <
d_c$ as a consequence of strong fluctuation effects.  Directly at the
critical dimension, one expects logarithmic corrections to the above
mean-field results.

The idea is now to combine several (say a number $k$) of such DP
processes, i.e., to consider the additional reactions
\begin{eqnarray}
\label{branchB}
&B \to B + B \quad &{\rm with \ rate \ } \sigma_B \ , \\
\label{decayB}
&B \to \emptyset \quad &{\rm with \ rate \ } \mu_B \ , \\
\label{coagulB}
&B + B \to B \quad &{\rm with \ rate \ } \lambda_B \ ,
\end{eqnarray}
for particle species $B,C,\ldots$, which are coupled {\em
unidirectionally} via the transformation reactions
\begin{eqnarray}
\label{trans}
&A \to B  \quad &{\rm with \ rate \ } \mu_{AB} \ , \\
\label{transA}
&A \to C  \quad &{\rm with \ rate \ } \mu_{AC} \ , \\
\label{transB}
&B \to C  \quad &{\rm with \ rate \ } \mu_{BC} \ ,
\end{eqnarray}
etc., but {\em without feedback}, i.e., we do not allow processes of
the type $B \to A$.  This prescription therefore defines a {\em
hierarchical structure} of coupled DP processes.  For simplicity, we
choose identical diffusion constants $D$ on each hierarchy level.

Without the coupling reactions (\ref{trans}), $\ldots$, for each
species of particles there is a continuous DP transition at $r_i = 0$,
$i=A,B,\ldots$.  But the situation changes in an interesting way when
the transformation processes are switched on (except for species $A$,
which is not influenced by what happens on the higher hierarchy
levels).  Let us first consider the simplest case of two particle
species ($k=2$) --- the generalization to further hierarchy levels
will then be straightforward.  The mean-field rate equation
(\ref{mfeq}) for species $A$ remains unchanged, albeit with a modified
parameter
\begin{equation}
\label{mfra}
r_A = (\mu_A + \mu_{AB} - \sigma_A) / D \ .
\end{equation}
Notice that for the first hierarchy level, the sole effect of the
transformation reactions is an increase of the total decay rate
$\mu_A^{\rm tot} = \mu_A + \sum_i \mu_{Ai}$.

The rate equation for species $B$, however, contains a new gain term
describing the feeding-in of particles via the reaction (\ref{trans}),
proportional to the density of $A$ particles present.  Thus one
obtains
\begin{eqnarray}
\label{mfeqB}
{\partial n_B(x,t) \over \partial t} = &&D \left( \nabla^2 - r_B
\right) n_B(x,t) \nonumber \\ &&- \lambda_B n_B(x,t)^2 + \mu_{AB}
n_A(x,t) \ ,
\end{eqnarray}
where
\begin{equation}
\label{mfrb}
r_B = (\mu_B - \sigma_B) / D \ .
\end{equation}
Notice that within the {\em mean-field} approximation, $n_A$ plainly
acts as an external source term.  Once fluctuations are important,
however, i.e., for $d < d_c = 4$, such a simple picture breaks down,
especially at the multicritical point to be discussed below, where the
averages and correlations of {\em both} $n_A(x,t)$ and $n_B(x,t)$ are
governed by power laws.

We may now again search for a stationary solution $n_B$ of
Eq.~(\ref{mfeqB}), as a function of the mean-field density $n_A$.  The
general solution of the ensuing quadratic equation is
\begin{equation}
\label{mfdenB}
n_B = \left[ \left( {D r_B \over 2 \lambda_B} \right)^2 + {\mu_{AB}
\over \lambda_B} \, n_A \right]^{1/2} - {D r_B \over 2 \lambda_B} \ .
\end{equation}
Thus, for $r_A > 0$, where $n_A = 0$, one finds $n_B = D (|r_B| - r_B)
/ 2 \lambda_B$, which is zero for $r_B > 0$, and becomes equal to $n_B
= D |r_B| / \lambda_B$ for $r_B < 0$.  When species $A$ is in the
inactive phase, the $A$ and $B$ hierarchy levels are effectively
decoupled, and we therefore expect an ordinary DP active / absorbing
transition for species $B$ at $r_B = 0$, as in the case $\mu_{AB} =
0$.

\begin{figure}
  \centerline{\epsfysize 5.5cm \epsfbox{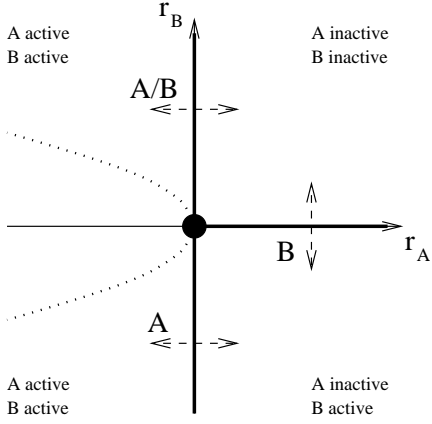}} \vskip 0.3 truecm
\caption{Mean-field phase diagram for the two-level coupled DP
  process.  The arrows mark DP active/absorbing transitions for the
  $A$ and $B$ particle species, respectively.  The dotted parabola
  denotes the boundary of the multicritical regime, which includes
  $|r_A| = |r_B| = |r| \to 0$.}
\label{phdiag}
\end{figure}

For $r_A < 0$, on the other hand, we have to insert (\ref{staden})
into Eq.~(\ref{mfdenB}).  One may now distinguish two situations: (i)
For $(D r_B / 2 \lambda_B)^2 \gg D |r_A| \mu_{AB} / \lambda_A
\lambda_B$, we can approximate
\begin{equation}
\label{apdenB}
n_B \approx {D |r_B| \over 2 \lambda_B} \left[ 1 + {D |r_A| \mu_{AB}
\over 2 \lambda_A \lambda_B} \left( {2 \lambda_B \over D r_B}
\right)^2 \right] - {D r_B \over 2 \lambda_B} \ ,
\end{equation}
and consequently for $r_B < 0$, we find that $n_B > 0$, and the $B$
species is in its active state --- the DP transition at the half line
$(r_A < 0, r_B = 0)$ present in the uncoupled system has disappeared
(see Fig.~\ref{phdiag}).  Instead, for $r_B > 0$ the terms
proportional to $r_B$ cancel in Eq.~(\ref{apdenB}), and
\begin{equation}
\label{denDPB}
n_B \approx |r_A| \mu_{AB} / \lambda_A r_B \ ,
\end{equation}
i.e., the density of species $B$ vanishes as the critical point $r_A =
0$ of species $A$ is approached, and with the mean-field DP exponent
$\beta = 1$.  Effectively, the DP critical half line $(r_A < 0, r_B =
0)$ for species $B$ has been rotated to $(r_A = 0, r_B > 0)$ in the
coupled system.  The location of the DP critical lines for both
species $A$ and $B$ is shown in the phase diagram of
Fig.~\ref{phdiag}, where the dotted parabola represents the boundary
curve separating the two different regimes for $r_A < 0$.  Notice also
that in this case the ``mass'' terms (the contributions linear in
$n_A$, $n_B$) in Eq.~(\ref{mfeqB}) vanish, and one expects both
spatial and temporal power-law correlations, obviously characterized
by $\nu_\perp = 1/2$ and $z = 2$.  We shall later see that indeed the
new critical half line $(r_A = 0, r_B > 0)$ for species $B$ is in the
DP university class.  Summarizing this regime, we may say that the $B$
particles are ``slaved'' by the behavior of the $A$ species.

(ii) The other regime, inside of the dotted parabola in
Fig.~\ref{phdiag}, is defined by the condition $(D r_B / 2
\lambda_B)^2 \ll D |r_A| \mu_{AB} / \lambda_A \lambda_B$, which
includes the special case when the critical points of both hierarchy
levels are approached uniformly, $|r_A| = |r_B| = |r| \to 0$.  Now we
can neglect the terms $\sim r_B$ in Eq.~(\ref{mfdenB}), which yields
\begin{equation}
\label{mcden}
n_B \approx \left( {D |r_A| \mu_{AB} \over \lambda_A \lambda_B}
\right)^{1/2} = \left( {\mu_{AB} \over \lambda_B} \, n_A \right)^{1/2}
\ .
\end{equation}
This implies that the density exponents on each hierarchy level $i$
are {\em different} in this regime,
\begin{equation}
\label{denexp}
n_i \approx |r_A|^{\beta_i} \ ,
\end{equation}
where $\beta_1 = \beta_A = 1$, and $\beta_2 = \beta_B = 1/2$ in the
mean-field approximation.  It is to be expected, however, that the
other independent scaling exponents $\nu_\perp$ and $z$ remain
unaltered; but of course the density decay exponent at the critical
point $\alpha_i$ will depend on the hierarchy level $i$, according to
Eq.~(\ref{scarel}),
\begin{equation}
\label{dendec}
n_i(t) \sim t^{-\alpha_i} \ , \quad \alpha_i \equiv \beta_i / z
\nu_\perp \ .
\end{equation}

Considering the two-species phase diagram (Fig.~\ref{phdiag}) again,
we see that {\em three} critical half lines converge at the special
point $r_A = r_B = 0$.  The new critical behavior in this regime can
therefore be interpreted as the effect of a {\em multicritical point}
\cite{ourprl}.  When this special point in the phase diagram is
approached along a line crossing the dotted parabola in
Fig.~\ref{phdiag}, one expects a {\em crossover} from ordinary DP to
the new multicritical behavior described by the density exponents
$\beta_i$ and $\alpha_i$.  For $|r_A| = |r_B| = |r| \to 0$, the
crossover features are all encoded in the generalized scaling function
\begin{equation}
\label{genscal}
n_B(r,\mu_{AB},x,t) = |r|^{\beta_1} {\hat n}_A(|r|^{-\phi} \mu_{AB} /
D, x/\xi_\perp, t/\xi_\parallel) \ ,
\end{equation}
where $\xi_\perp \sim |r|^{-\nu_\perp}$ and $\xi_\parallel \sim
|r|^{-\nu_\parallel}$ as in DP.  This defines the {\em crossover
exponent} $\phi$, which constitutes a new scaling exponent associated
with the coupling $\mu_{AB}$.  Comparing with Eq.~(\ref{mfdenB}), we
identify
\begin{equation}
\label{mfphi}
\phi = 1
\end{equation}
within the mean-field approximation.  Furthermore, we can use the
above mean-field results, which, at the multicritical point, imply
that ${\hat n}(y,0,0) \sim y^{1/2}$ for $y \to \infty$. Consequently
we have
\begin{equation}
\label{scamc}
\beta_2 = \beta_1 - \phi / 2 \ ,
\end{equation}
which relates the new density exponent $\beta_2$ to the independent
crossover exponent $\phi$.  Of course, Eq.~(\ref{scamc}) is satisfied
by the mean-field values.  In the fluctuation-dominated regime $d <
d_c = 4$, however, there will in general be $O(\epsilon = 4-d)$
corrections to both the critical exponents {\em and} the scaling
functions, which will in turn lead to a modification of the scaling
relation (\ref{scamc}).

One may now readily generalize to higher hierarchy levels.  For
example, for $k=3$, one finds the same rate equations (\ref{mfeq}) and
(\ref{mfeqB}) for species $A$ and $B$, respectively, but where now
$r_A = (\mu_A + \mu_{AB} + \mu_{AC} - \sigma_A) / D$, and $r_B =
(\mu_B + \mu_{BC} - \sigma_B) / D$.  The mean-field equation for
$n_C(x,t)$ reads
\begin{eqnarray}
\label{mfeqC}
&&{\partial n_C(x,t) \over \partial t} = D \left( \nabla^2 - r_C \right)
  n_C(x,t) \nonumber \\
&&\quad - \lambda_C n_C(x,t)^2 + \mu_{BC} n_B(x,t) + \mu_{AC} n_A(x,t) \
,
\end{eqnarray}
which has the general stationary solution
\begin{equation}
\label{mfdenC}
n_C = \left[ \left( {D r_C \over 2 \lambda_C} \right)^2 + {\mu_{BC}
\over \lambda_C} \, n_B + {\mu_{AC} \over \lambda_C} \, n_A
\right]^{1/2} - {D r_C \over 2 \lambda_C} \ .
\end{equation}
A detailed analysis then reveals that, in analogy with the two-level
hierarchy, there are regions in phase space where the $C$ species
evolves independently of the lower hierarchy levels. On the other hand
other regimes exist where $n_C$ is slaved by either $n_A$ or $n_B$. In
addition, as before, a new DP transition may arise for the $C$
particles under appropriate conditions, when $r_A \to 0$ or $r_B \to
0$.  Furthermore, the previous $k=2$ multicritical regime occurs when
either $r_A > 0$, and $|r_B| = |r_C| \to 0$ simultaneously, or $r_B >
0$, and $|r_A| = |r_C| \to 0$.

As all these features are already contained in our above investigation
of the two-level coupled DP process, we restrict ourselves to the new
behavior emerging for $k = 3$, when {\em all three} critical points
coincide, i.e., $|r_A| = |r_B| = |r_C| = |r| \to 0$.  More generally,
in mean-field theory this multicritical regime (with altogether {\em
seven} critical quarter planes merging at $r=0$) is characterized by
the conditions $r_A < 0$, $(D r_B / 2 \lambda_B)^2 \ll D |r_A|
\mu_{AB} / \lambda_A \lambda_B$ (as for $k=2$), and $(D r_C / 2
\lambda_C)^2 \ll \mu_{BC} (D |r_A| \mu_{AB} / \lambda_A
\lambda_B)^{1/2} / \lambda_C + D |r_A| \mu_{AC} / \lambda_A
\lambda_C$.  At this special point in parameter space, $n_A$ and $n_B$
vanish as in Eqs.~(\ref{staden}) and (\ref{mcden}), respectively,
while
\begin{equation}
\label{mcdenC}
n_C \approx \left( {D |r| \mu_{AB} \mu_{BC}^2 \over \lambda_A
\lambda_B \lambda_C^2} \right)^{1/4} = \left( {\mu_{BC} \over
\lambda_C} \, n_B \right)^{1/2} \ .
\end{equation}
Therefore $\beta_3 = \beta_C = 1/4$ in the mean-field approximation.
Notice also that the indirect $A \to C$ transformation rate $\mu_{AC}$
does {\em not} enter this expression (\ref{mcdenC}), which only
depends on the coupling rates $\mu_{AB}$ and $\mu_{BC}$ between {\em
adjacent} hierarchy levels. As the latter are not at all influenced by
the presence of lower levels, this would suggest that there exists
{\em only one} crossover exponent $\phi$ describing all the
multicritical points generated by the unidirectional coupling of the
DP processes (with $\phi = 1$ in mean-field theory).

For the density exponents, we conclude that within the mean-field
approximation on hierarchy level $i$, using $\nu_\parallel \equiv z
\nu_\perp = 1$,
\begin{equation}
\label{denli}
\alpha_i = \beta_i = 1 / 2^{i-1} \ .
\end{equation}
This result should describe the coupled DP multicritical point
quantitatively correctly for spatial dimensions $d > d_c = 4$.
Furthermore, on the mean-field level the scaling relation
(\ref{scamc}) generalizes to
\begin{equation}
\label{mcscai}
\beta_i = \beta_{i-1} - \phi / 2^{i-1} = \beta_1 - \phi (1 - 1 /
2^{i-1}) \ ,
\end{equation}
compare Eq.~(\ref{mcdenC}).  Here the observation that only the direct
transformation rates between neighboring hierarchy levels affect the
leading contribution has entered crucially.  Again, the mean-field
results (\ref{denli}) and (\ref{mfphi}) satisfy Eq.~(\ref{mcscai})
trivially.  However, as noted above, the scaling relation
(\ref{mcscai}) will be modified below the upper critical dimension
$d_c = 4$, as a consequence of $O(\epsilon = 4 - d)$ corrections to
the scaling function for the equations of state, $n_i(r)$.

\section{Renormalization Group Calculations}
\label{renorm}


\subsection{Preliminaries}
\label{prelim}


We now turn to a detailed presentation of our field-theoretic 
calculations.  As we have pointed out in the previous sections, the 
effects of fluctuations invalidate a simple mean-field approach for 
dimensions $d < d_c = 4$.  For that reason we will employ 
field-theoretic renormalization group methods, which allow both for a 
proper derivation of scaling, as well as a systematic 
$\epsilon$-expansion calculation of critical exponents, below the 
upper critical dimension $d_c = 4$.

Our starting point for a systematic treatment of the coupled-DP
reaction scheme given by (\ref{branch})-(\ref{coagul}),
(\ref{branchB})-(\ref{trans}) is an appropriate master equation. 
On a microscopic level this comprises an exact description of the 
dynamics.  From this equation it is then a straightforward process to 
derive an effective field theory: First the master equation is mapped 
onto a second-quantized bosonic operator representation, which is in
turn mapped onto a bosonic field theory.  This procedure is now 
standard, and we refer to Ref.~\cite{masfth} for further details.  In 
our case, for the two-species coupled DP system we end up with the 
following action
\begin{eqnarray}
\label{action0}
& & S = \int \! d^dx \int \! dt \,\bigl\{ {\bar a} \left[ \partial_t 
+ D (r_A - \nabla^2) \right] a - \sigma_A {\bar a}^2 a + \nonumber \\ 
& & \quad\qquad\qquad + \lambda_A ( {\bar a} a^2 + {\bar a}^2 a^2 )
- \mu_{AB} {\bar b}a+  \\ 
& & {\quad +\bar b} \left[ \partial_t + D(r_B-\nabla^2) \right] b - 
\sigma_B {\bar b}^2 b + \lambda_B ( {\bar b} b^2 + {\bar b}^2 b^2)
\bigr\} \ , 
\nonumber
\end{eqnarray}
where we have omitted terms related to the initial state. Aside from 
the taking of the continuum limit, the derivation of this action is 
{\em exact}, and in particular no assumptions regarding the precise 
form of the noise are required.  Note that if we neglect the terms 
in the action (\ref{action0}) quadratic in the response fields 
$\bar a$, $\bar b$, then we recover a description identical to the
mean-field equations (\ref{mfeq}) and (\ref{mfeqB}), provided we
associate the fields $a({\bf x},t$), $b({\bf x},t)$ with the 
coarse-grained local densities of the $A$, $B$ particles. In general, 
however, below the upper critical dimension $d_c=4$, the terms 
quadratic in $\bar a$, $\bar b$ (corresponding to noise in a Langevin 
description) cannot be neglected.

It is now convenient to rescale the fields according to ${\bar a} =
(\lambda_A / \sigma_A)^{1/2} {\bar \psi}_0$, $a = (\sigma_A / 
\lambda_A)^{1/2} \psi_0$, ${\bar b} = (\lambda_B / \sigma_B)^{1/2} 
{\bar \varphi}_0$, $b = (\sigma_B / \lambda_B)^{1/2} \varphi_0$, and 
also define new couplings $u_0 = 2 (\sigma_A \lambda_A)^{1/2}$, 
$u_0' = 2 (\sigma_B \lambda_B)^{1/2}$, as well as $\mu_0 = \mu_{AB} 
(\sigma_A \lambda_B / \sigma_B \lambda_A)^{1/2}$ (henceforth, the
subscript ``0'' denotes unrenormalized quantities).  If we introduce 
a length scale $\kappa^{-1}$ and correspondingly measure times in 
units of $\kappa^{-2}$ (i.e., $[D_0] = \kappa^0$), we find that the 
new fields have scaling dimension $\kappa^{d/2}$, while $[r_A] = 
[r_B] = [\mu_0] = \kappa^2$, which are thus relevant perturbations 
in the RG sense. On the other hand, $[u_0] = [u_0'] = 
\kappa^{2-d/2}$, and the corresponding DP non-linearities become
marginal in $d_c = 4$ dimensions, as expected \cite{regfth}.  It is 
important to note, however, that $[\lambda_A] = [\lambda_B] = 
\kappa^{2-d}$, and hence these couplings are {\em irrelevant} as 
compared to $u_0$ and $u_0'$, and may be omitted in the effective 
action.  Finally, we set $u_0' = u_0$, such that the theory remains
renormalizable with equal diffusion constants \cite{referee}, and 
consequently arrive at
\begin{eqnarray}
\label{action1}
S_{\rm eff} & & = \int \! d^dx \int \! dt \, \Bigl\{ {\bar \psi}_0 
\left[ \partial_t + D_0 (r_A - \nabla^2) \right] \psi_0 - \nonumber \\
& & \qquad\quad - {u_0 \over 2} \left( {\bar \psi}_0^2 \psi_0 
- {\bar \psi}_0 \psi_0^2 \right) 
- \mu_0 {\bar \varphi}_0 \psi_0 + \label{effact} \\
& & + {\bar \varphi}_0 \left[ \partial_t + D_0 (r_B - \nabla^2) 
\right] \varphi_0 - {u_0 \over 2} \left( {\bar \varphi}_0^2 \varphi_0 
- {\bar \varphi}_0 \varphi_0^2 \right) \Bigr\} \ . \nonumber
\end{eqnarray}
We remark that this action is equivalent to the following set of coupled
Langevin equations
\begin{eqnarray}
&&\partial_t \psi_0 = D_0 (\nabla^2 - r_A) \psi_0 - {u_0 \over 2} 
\psi_0^2 + \zeta \ , \label{lanpsi} \\
&&\partial_t \varphi_0 = D_0 (\nabla^2 - r_B) \varphi_0 - {u_0 \over 2} 
\varphi_0^2 + \mu_0 \psi_0 + \eta \ , \label{lanphi}
\end{eqnarray}
which represent the obvious and expected generalizations of the 
mean-field equations (\ref{mfeq}) and (\ref{mfeqB}), with the 
multiplicative Langevin noise terms
\begin{eqnarray}
&&\langle \zeta(x,t) \zeta(x',t') \rangle = u_0 \psi_0 (x,t) \, \delta^d(x-x')
\delta(t-t') \ , \label{noipsi} \\
&&\langle \eta(x,t) \eta(x',t') \rangle = u_0 \varphi_0 (x,t) \, \delta^d(x-x')
\delta(t-t') \ . \label{noiphi}
\end{eqnarray}

Furthermore, for most of the analysis we will put $r_A = r_B = r_0$. 
There are several ways in which the effective action (\ref{action1}) 
can be studied.  We will begin by performing our analysis in the 
{\em inactive} phase, and postpone other methods (including 
diagonalization and active phase computations) until later on.  
Inspection of the above action (\ref{action1}) reveals that, as 
expected, the terms involving only the $\psi$ and $\bar\psi$ fields 
are exactly the same as in the well-known field theory for directed 
percolation (Reggeon field theory) \cite{regfth}, and their 
renormalization is entirely unaffected by the presence of the 
$\varphi$ and $\bar\varphi$ fields.  Hence we will begin by briefly 
reviewing the analysis of Reggeon field theory (RFT) in the inactive 
phase.

\subsection{DP field theory: Inactive phase calculation}
\label{DPinactivephase}

\begin{figure}
  \centerline{\epsfxsize 8.5cm \epsfbox{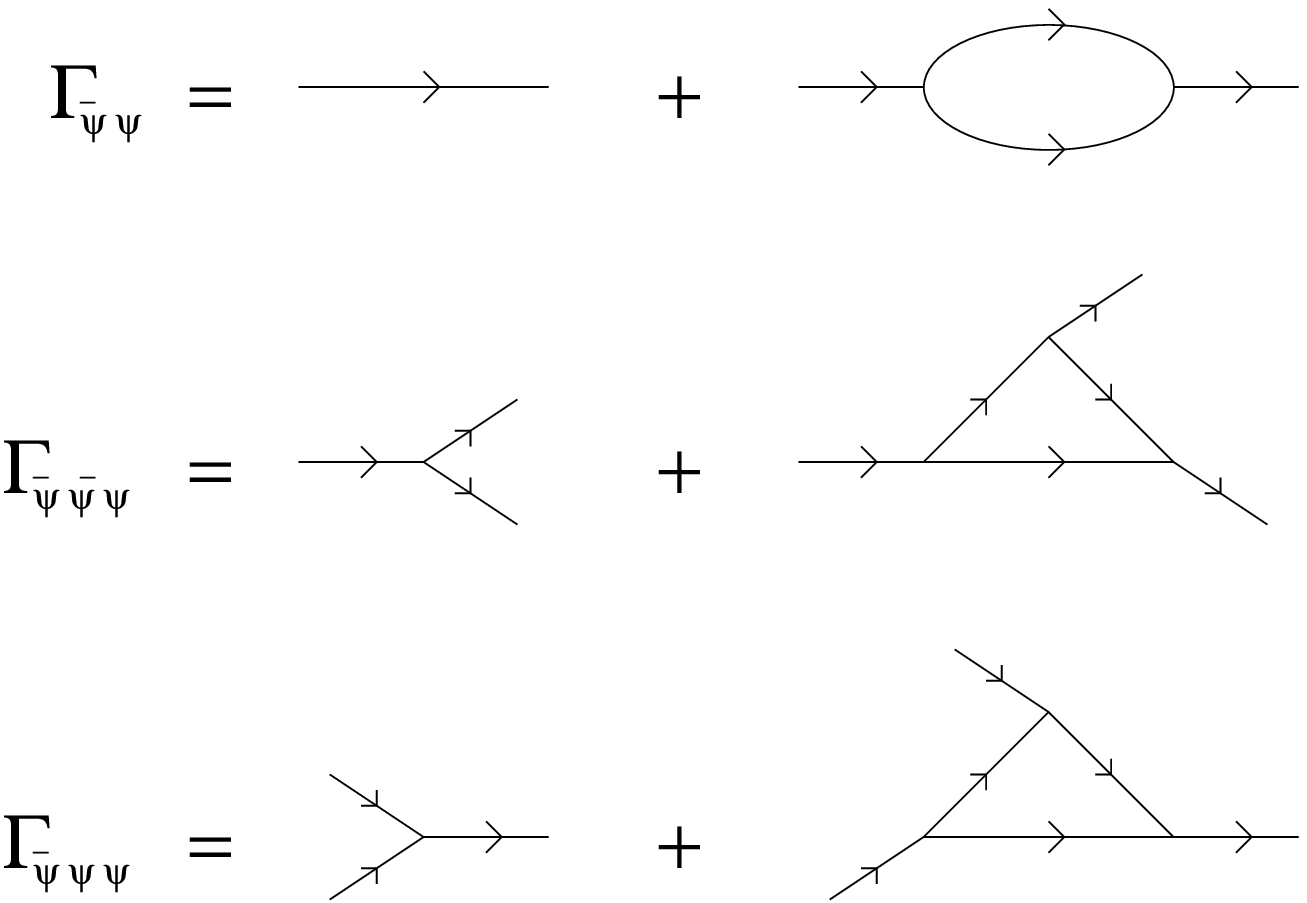}} \vspace{3mm}
\caption{Diagrams for the two- and three-point vertex functions to
  one-loop order for the pure DP field theory.}
\label{rftrenormdiag}
\end{figure}

The renormalization of the RFT action is very well known (see 
Ref.~\cite{rftren}). The renormalized parameters are defined as 
follows: 
\begin{eqnarray}
\label{rftrenorm}
& & \psi=Z_{\psi}^{1/2}\psi_0 \ , \quad 
\bar\psi=Z_{\psi}^{1/2}\bar\psi_0 \, \quad 
\tau=Z_{\tau}\tau_0\kappa^{-2} \ , \nonumber \\
& & D=Z_D D_0 \ , \quad u=Z_u u_0 A_d^{1/2}\kappa^{-\epsilon/2} \ ,
\end{eqnarray}
with $\epsilon=4-d$, $A_d=\Gamma(3-d/2)/2^{d-1}\pi^{d/2}$, and
$\tau_0=r_0-r_{0c}$, where $r_{0c}$ is the fluctuation-induced shift 
of the critical point.  From the diagrams for the two- and 
three-point vertex functions (see Fig.~\ref{rftrenormdiag}), we can 
determine the one-loop renormalized $Z$ factors.  Using dimensional 
regularization and a minimal subtraction scheme, the results are
\begin{eqnarray}
\label{Zfactorpsi}
& & Z_{\psi}=1 - {u_0^2 \over 8 D_0^2} 
{A_d\kappa^{-\epsilon}\over\epsilon} \ , \\
\label{ZfactorD}
& & Z_{D}=1+{u_0^2 \over 16 D_0^2}
{A_d\kappa^{-\epsilon}\over\epsilon} \ , \\
\label{Zfactortau}
& & Z_{\tau}=1-{3 u_0^2 \over 16 D_0^2}
{A_d\kappa^{-\epsilon}\over\epsilon} \ , \\
\label{Zfactoru}
& & Z_{u}=1-{5 u_0^2\over 16 D_0^2}
{A_d\kappa^{-\epsilon}\over\epsilon} \ ,
\end{eqnarray}
with $r_{0c}$ given by the recursive equation
\begin{equation}
\label{r0cshift}
r_{0c}=-{u_0^2\over 4 D_0^2} \int_p {1\over r_{0c} + p^2} \ ,
\end{equation}
where we have used the abbreviation 
$\int_p \ldots =\int \ldots d^dp/(2\pi)^d$.  Defining the flow 
functions $\zeta_{u}=\kappa\partial_{\kappa}\ln(u/u_0)$ etc., and 
with an effective coupling $v=u^2/16D^2$, the RG $\beta$ function 
has the form
\begin{equation}
\label{betav}
\beta_v=\kappa\partial_{\kappa}v=2v(\zeta_u-\zeta_D) = 
v ( -\epsilon+12 v) \ ,
\end{equation}
giving a stable, non-trivial fixed point 
$v^*=\epsilon/12+O(\epsilon^2)$.

The appropriate renormalization group equation for the renormalized
field $\langle\psi_R\rangle$, where the angular brackets denote 
averaging with respect to the RFT action, is
\begin{eqnarray}
\label{RFTrenormgpeq}
& & \left(\kappa{\partial\over\partial\kappa}+\zeta_{\tau}\tau
{\partial\over\partial\tau}+\zeta_D D{\partial\over\partial D}+
\zeta_v v{\partial\over\partial v}-{1\over 2}\zeta_{\psi}\right) 
\nonumber \\
& & \qquad\qquad\qquad\qquad\times\langle\psi_R(\kappa,\tau,D,v,x,t) 
\rangle=0 \ .
\end{eqnarray}
Defining the dimensionless field $\hat\psi$ as
\begin{equation}
\label{psidimless}
\langle\psi_R(\kappa,\tau,D,v,x,t)\rangle=\kappa^{d/2}
\hat\psi(\tau,v,\kappa x, \kappa^2 Dt) \ ,
\end{equation}
the solution of (\ref{RFTrenormgpeq}), obtained by means of the method
of characteristics $\kappa \to \kappa \ell$, when the coupling $v$ has 
run to its fixed point value $v^*$ is
\begin{eqnarray}
\label{RGpsisolution}
& & \langle\psi_R(\kappa,\tau,D,v,x,t)\rangle=\kappa^{d/2}
\ell^{(d-\zeta_{\psi}^*)/2} \nonumber \\
& & \qquad\qquad \times\hat\psi(\tau\ell^{\zeta_{\tau}^*},v^*,\kappa x
\ell,\kappa^2 Dt \ell^{2+\zeta_D^*}) \ .
\end{eqnarray}
By inserting the matching condition $\ell=|\tau|^{-1/\zeta_{\tau}^*}$,
we can now {\em derive} the scaling relation (\ref{scaling}) quoted in 
the previous section.  At the fixed point, we obtain
\begin{equation}
\langle\psi_R(\kappa,\tau,D,v,x,t)\rangle = |\tau|^\beta 
\hat\psi\left(v^*,{\kappa x \over |\tau|^{-\nu_{\perp}}},
{\kappa^2 Dt \over |\tau|^{-\nu_{\parallel}}}\right) \ ,
\end{equation}
where the exponents can be identified as combinations of the $\zeta$
functions evaluated at the non-trivial fixed point:
\begin{eqnarray}
\label{rftexponents}
& & \eta_{\perp}=-\zeta_D^*-\zeta_{\psi}^*=
-{\epsilon \over 12}+O(\epsilon^2) \ , \\
\label{DPeta}
& & \nu_{\perp}={1\over -\zeta_{\tau}^*}=
{1\over 2}+{\epsilon\over 16}+O(\epsilon^2) \ , \\
\label{DPnuperp}
& & \nu_{\parallel}={2+\zeta_D^*\over -\zeta_{\tau}^*}=
1+{\epsilon \over 12}+O(\epsilon^2) \ , \\
\label{DPnupar}
& & z={\nu_{\parallel}\over\nu_{\perp}}=
2+\zeta_D^*=2-{\epsilon\over 12}+ O(\epsilon^2) \ , \\
\label{DPz}
& & \beta={d-\zeta_{\psi}^*\over -2\zeta_{\tau}^*} = 
{\nu_{\perp}\over 2}(d+z-2+\eta_{\perp}) \nonumber \\
& & \quad = 1-{\epsilon\over 6}+O(\epsilon^2) \ .
\label{DPbeta}
\end{eqnarray}

Directly at the upper critical dimension $d_c = 4$ ($\epsilon = 0$), 
the power laws with these critical exponents are replaced with 
logarithmic corrections to the mean-field results $\eta_\perp = 0$, 
$\nu_\perp = 1/2$, $\nu_\parallel = 1$, $z = 2$, and $\beta = 1$ 
(see also Ref.~\cite{crslp2}).  The flow equation for $v(\ell)$ 
becomes $\ell\,dv(\ell)/d\ell = \beta_v(\ell)=12v(\ell)^2$, which is
solved by $v(\ell) = v [1-12v \ln \ell]^{-1}$, where $v = v(\ell=1)$.
Similarly, $\ell\,dD(\ell)/d\ell = \zeta_D(\ell) D(\ell) = - v 
D(\ell) [1-12v \ln \ell]^{-1}$ has the solution $D(\ell) = D[1-12v 
\ln \ell]^{1/12}$, and $\ell\,d\tau(\ell)/d\ell = \zeta_\tau(\ell) 
\tau(\ell) = (-2+3v [1-12v \ln \ell]^{-1}) \tau(\ell)$ is solved by 
$\tau(\ell) = \tau \ell^{-2}[1-12v \ln \ell]^{-1/4}$. We now integrate 
the flow equations until $|\tau(\ell)|=1$, or $\ell \sim 
|\tau|^{1/2}(-\ln |\tau|)^{-1/8}$. This yields immediately the
divergence of both the correlation length $\xi_\perp$ and the 
characteristic time scale $\xi_\parallel$ upon approaching the phase 
transition,
\begin{eqnarray}
& & \xi_\perp \sim \ell^{-1} \approx 
|\tau|^{-1/2}(-\ln |\tau|)^{1/8} \ , \\
\label{xilog}
& & \xi_\parallel \sim \ell^{-2} D(\ell)^{-1} \approx 
|\tau|^{-1}(-\ln |\tau|)^{1/6} \ . 
\label{tclog}
\end{eqnarray}
In order to obtain the corresponding logarithmic correction for the
density exponent $\beta$, we employ the solution (\ref{RGpsisolution}) 
of the RG equation and the mean-field result (\ref{staden}), and find
\begin{equation}
\langle\psi_R(\kappa,\tau,D,v)\rangle \sim \kappa^{d/2} \ell^{d/2} 
C(\ell)^{-1/2} {|\tau(\ell)| \over v(\ell)^{1/2}} \ ,
\label{opflow}
\end{equation}
where $C(\ell) = \exp(\int_1^\ell \zeta_\psi(\ell') d\ell'/\ell')$, or
equivalently, $\ell dC(\ell)/d\ell = \zeta_\psi(\ell) C(\ell) = 2 v
C(\ell) [1-12v \ln \ell]^{-1}$, with the solution $C(\ell) = [1-12v 
\ln \ell]^{-1/6}$.  Combining everything, and setting $d=4$ in 
Eq.~(\ref{opflow}) finally yields
\begin{equation}
\langle\psi_R\rangle \sim |\tau| (-\ln |\tau|)^{1/3} \ .
\label{oplog}
\end{equation}
Notice that we had to take care and keep track of the dangerous
irrelevant variable $v$ here.

\subsection{Coupled DP field theory: Inactive phase}
\label{coupleDPinactivephase}

\begin{figure}
  \centerline{\epsfxsize 8.5cm \epsfbox{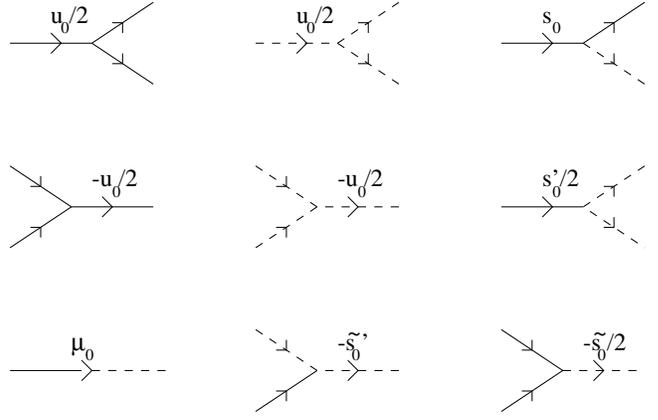}} \vspace{3mm}
\caption{The vertices of the full action $S_{\rm mc}$.}
\label{fullvertices}
\end{figure}

We now return to the renormalization of the coupled DP field theory. 
Right from the outset we must take into account one key feature of the 
full theory --- namely that, on physical grounds, one expects the generation 
of {\em additional} mixed cubic vertices. Physically, these novel vertices 
correspond to the additionally generated processes $A \to A + B$, 
$A \to B + B$, $A + A \to B$, and $A + B \to A$, with rates $\sigma_{AB}$, 
$\sigma_{AB}'$, 
$\lambda_{AB}$, and $\lambda_{AB}'$, say. These vertices must be introduced 
from the very beginning, and hence we have to replace the above action 
(\ref{action1}) with $S_{\rm mc}=S_{\rm eff}+\Delta S$, where
\begin{eqnarray}
\label{deltaS}
& & \Delta S=\int \! d^dx \int \! dt \, \left[ - s_0 \bar\varphi_0
\bar\psi_0 \psi_0 - {s_0'\over 2} {\bar\varphi_0}^2\psi_0 \right. 
\nonumber \\
& & \left. \qquad\qquad\qquad\qquad + {\tilde s_0 \over 2} 
\bar\varphi_0 \psi_0^2 + \tilde s_0' \bar\varphi_0 \varphi_0 \psi_0
\right] \ . 
\end{eqnarray}
Note that in the shifted theory used
above, the new reaction processes also modify the bare parameters 
$r_A$, $\mu_0$, and 
$u_0$, and furthermore lead to the identification $s_0 \sim 
\sigma_{AB} \geq 0$, $s_0' \sim \sigma_{AB}' \geq 0$, $\tilde s_0 \sim 
- \lambda_{AB} \leq 0$, and $\tilde s_0' \sim \lambda_{AB}' \geq 0$.  In 
the effective Langevin-type description, Eqs.~(\ref{lanpsi})-(\ref{noiphi}) 
are then replaced by
\begin{eqnarray}
\partial_t \psi_0 = D_0 (\nabla^2 - r_A) \psi_0 - {u_0 \over 2}
\psi_0^2 + \zeta \ , \label{lanpsim} \\
\partial_t \varphi_0 = D_0 (\nabla^2 - r_B) \varphi_0 - {u_0 \over
  2} \varphi_0^2 \ \ \ \ \nonumber \\ 
- {\tilde s_0 \over 2} \psi_0^2 - \tilde s_0' \psi_0 
\varphi_0 + \mu_0 \psi_0 + \eta \ , \label{lanphim}
\end{eqnarray}
where the noise terms satisfy
\begin{eqnarray}
\zeta(x,t)  &=& a \xi_1(x,t), \label{noipsim} \\
\eta(x,t) &=& b \xi_2(x,t) + c \xi_1(x,t) \ .
 \label{noiphim}
\end{eqnarray}
Here $\xi_1$ and $\xi_2$ are uncorrelated white noise variables of variance
one, and the coefficients $a$, $b$ and $c$ satisfy:
\begin{eqnarray}
a^2&=&u_0 \psi_0, \\
b^2&=&u_0 \varphi_0 +\left(s_0'-{s_0^2 \over u_0}\right)\psi_0, \\
c^2&=&{s_0^2 \over u_0}\psi_0.
\label{coeffnoise}
\end{eqnarray}
The complete vertices of the full action $S_{\rm mc}$ are depicted in
Fig.~\ref{fullvertices}.
The propagators for the $\psi$ and $\varphi$ fields are denoted by 
solid and dashed lines, respectively. The next task is to compute the 
renormalized couplings and RG fixed points of the full theory.  To 
begin with, we notice that the DP couplings in the action 
(\ref{action1}) are renormalized in the same way as in RFT, i.e., the 
factors $Z_{\psi}=Z_{\varphi}$, $Z_{\tau}$, $Z_D$, and $Z_u$ are 
{\em identical} with those in Eqs.~(\ref{Zfactorpsi})--(\ref{Zfactoru}).  
Hence we can conclude that the stable fixed points for the dimensionless 
renormalized coupling $u=Z_u u_0 A_d^{1/2}\kappa^{-\epsilon/2}$ reads
as in DP:
\begin{equation}
\label{u&u'fp}
v^*=[(u/4D)^*]^2 = \epsilon/12+O(\epsilon^2) \ .
\end{equation}
For this reason, the critical exponents $\eta_{\perp}$, $\nu_{\perp}$
and $z$ remain those of the DP universality class for the second 
hierarchy level (i.e., for the $B$ species), and in fact for higher 
hierarchy levels as well.

\begin{figure}
  \centerline{\epsfxsize 8.5cm \epsfbox{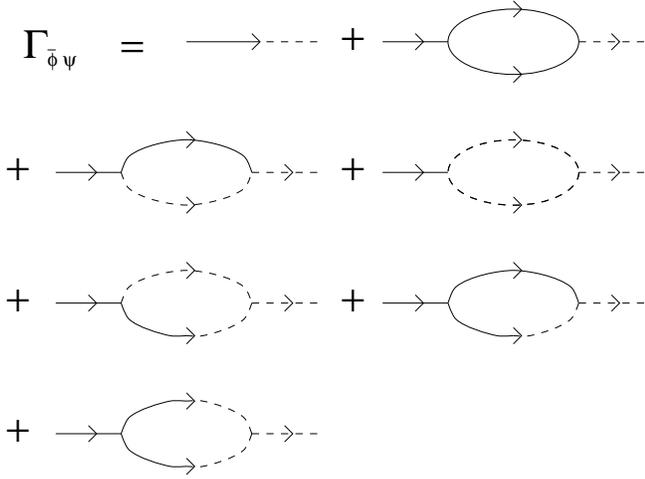}} \vspace{3mm}
\caption{
\label{MixedVertexFunctions}
``Mixed'' two-point vertex function $\Gamma_{{\bar \varphi} \psi}$ to
one-loop order for the coupled DP field theory.}
\label{murenormdiag}
\end{figure}

However, the same will {\em not} be true for the exponents $\beta_i$
(for $i>1$). For example, the exponent $\beta_2$ {\em is} affected by 
the renormalization of $\mu_0$, and hence only the exponent $\beta_1$ 
will remain the same as in DP.  In order to compute the renormalization 
of $\mu_0$, we must consider the diagrams renormalizing the ``mixed'' 
two-point vertex function $\Gamma_{\bar\varphi\psi}$, as depicted in 
Fig.~\ref{murenormdiag}.  At the normalization point $q=\omega=0$, 
$\tau=1$, we find
\begin{eqnarray}
\label{murenormstep1}
& & \Gamma_{\bar\varphi\psi}^{\rm NP} =-\mu_0 \left[ 1 - {u_0 (s_0 + 
\tilde s_0') \over 4D_0^2} \int_p {1\over (\kappa^2+p^2)^2} \right. 
\nonumber \\
& & \qquad\qquad\qquad\qquad - {u_0^2 \mu_0 \over 8 D_0^3} \int_p
{1\over (\kappa^2+p^2)^3} \nonumber \\
& & \qquad\qquad \left. - {2s_0 \tilde s_0' + u_0 (s_0' + \tilde s_0) 
\over 4D_0 \mu_0} \int_p {1\over \kappa^2+p^2} \right] \ . 
\end{eqnarray}
Notice that the integral in the last term of (\ref{murenormstep1})
diverges in $d=2$.  In the same way as in the shift of the percolation 
threshold in DP, see Eq.~(\ref{r0cshift}), we may take care of this 
UV divergence by means of an additive renormalization, and then 
multiplicatively renormalize the UV poles in $d = 4$.  Thus, defining 
the dimensionless renormalized coupling
\begin{equation}
\label{murdef}
\mu = Z_{\mu} (\mu_0 - \mu_{0c}) \kappa^{-2} \ ,
\end{equation}
with the associated flow function $\zeta_{\mu}=\kappa\partial_{\kappa}
\ln(\mu/\mu_0)$, we have
\begin{equation}
\label{mucshift}
\mu_{0c} = {2s_0\tilde s_0' + u_0 (s_0' + \tilde s_0) \over 4D_0}
\int_p {1\over \kappa^2+p^2} \ ,
\end{equation}
and
\begin{eqnarray}
\label{murenprod}
(Z_{\psi} Z_{\varphi})^{1/2} Z_{\mu} &=& 1 - {u_0 (s_0 + \tilde s_0')
\over 4D_0^2} \int_p {1\over (\kappa^2+p^2)^2} \nonumber \\
&&\qquad - {u_0^2 \mu_0 \over 8 D_0^3} \int_p {1\over (\kappa^2+p^2)^3} 
\ .
\end{eqnarray}
We will later see that the prefactor multiplying the shift $\mu_{0c}$ 
in Eq.~(\ref{mucshift}), which involves the various mixed three-point
couplings, actually vanishes in an appropriate parameter subspace 
containing {\em both} emerging fixed lines, see below.  In principle,
additional additive renormalizations would be required to render
$\partial_{q^2} \Gamma_{\bar\varphi\psi}^{\rm NP}$ and 
$\partial_\omega \Gamma_{\bar\varphi\psi}^{\rm NP}$ UV-finite 
\cite{referee}.  These counterterms, to be added to the action 
(\ref{action1}), would be of the form
\begin{equation}
\label{counteradd}
\int d^dx \int dt \, \bar{\varphi} \left( A \partial_t - B \nabla^2 \right) 
\psi \ .
\end{equation}  
However, as both $A$ and $B$ are again proportional to the prefactor of
the integral in Eq.~(\ref{mucshift}), they all vanish at the fixed lines 
to be discussed later.
A subtle point which can be raised concerning these counterterms is the 
stability of the scaling behavior of the theory (in other words the 
stability of the fixed lines to be derived later on) against the 
introduction of a term like Eq.~(\ref{counteradd}) into the original 
action (\ref{action1}). After 
this paper was submitted for publication, Janssen \cite{janssen2} has 
shown that indeed the scaling behavior is unaffected by the introduction 
of such terms, and thus $\mu_0$ is the only mixed coupling constant that 
needs to be introduced. 

Note also that the final diagram in 
Fig.~{\ref{murenormdiag} [see the second lines of 
Eqs.~(\ref{murenormstep1}) and (\ref{murenprod})] is UV-finite in $d=4$, 
and so for the moment we shall neglect it in a {\em minimal} subtraction 
scheme (this is, however, a somewhat subtle point in the active phase, 
which will be discussed in more detail in Sec.~\ref{techdiff}).  
Certainly, for $\mu \ll 1$, i.e., in an additional expansion in the 
transmutation rate $\mu$, this diagram is suppressed as compared to the 
other contributions.  We then find
\begin{equation}
\label{zetmu}
Z_\mu = 1 + \left( {u_0^2 \over 8 D_0^2} - {u_0 (s_0 + \tilde s_0') 
\over 4 D_0^2} \right) {A_d \kappa^{-\epsilon} \over \epsilon} \ ,
\end{equation}
and after defining $g=s/D$ and $\tilde g'=\tilde s'/D$, we have
\begin{eqnarray}
\label{zetamu}
\zeta_\mu &=& - 2 - 2 v + \sqrt{v} (g + \tilde g') \nonumber \\
&=& - 2 - {\epsilon \over 6} + {1 \over 2} \sqrt{\epsilon \over 3} 
\left( g^* + \tilde g'^* \right) \ ,
\end{eqnarray}
where in the second line we have inserted the DP fixed point
(\ref{u&u'fp}).

An inspection of Eq.~(\ref{murenormstep1}) now shows that, in order 
to compute the renormalization of $\mu_0$, we must first consider the
renormalization of the various mixed three-point couplings,
\begin{eqnarray}
\label{srenormdef}
& & s=Z_{s} s_0 A_d^{1/2}\kappa^{-\epsilon/2} \ , \quad
s'=Z_{s'} s'_0 A_d^{1/2}\kappa^{-\epsilon/2} \ , \nonumber \\
& & \tilde s=Z_{\tilde s}\tilde s_0 A_d^{1/2}\kappa^{-\epsilon/2} \ ,
\quad \tilde s'=Z_{\tilde s'}\tilde s'_0 A_d^{1/2}\kappa^{-\epsilon/2} 
\ .
\end{eqnarray}
Evaluating first the renormalization of $s'$, which can be calculated
from the diagrams shown in Fig.~\ref{genrenormdiag1}(a), we find
\begin{eqnarray}
\label{sgammarenorm}
& & \Gamma_{\bar\varphi\bar\varphi\psi_R}^{NP}={-s' A_d^{-1/2} 
\kappa^{\epsilon/2} \over Z_{\psi}^{1/2} Z_{\varphi}Z_{s'}} \left[ 1 
- \left( {u_0 s_0 \tilde s_0\over s_0'}+u_0 s_0 + u_0^2 \right.\right.
\nonumber \\
& & \left.\left. + u_0 \tilde s_0' + {u_0 s_0 \tilde s_0' \over s_0'}
+ {s_0^2 \tilde s_0' \over s_0'} \right) {1\over 2D_0^2} \int_p
{1\over (\kappa^2+p^2)^2}\right] \ .
\end{eqnarray}
Inserting the appropriate expressions for $Z_{\psi}$ and $Z_{\varphi}$
from Eq.~(\ref{Zfactorpsi}) and using $\int_p(\kappa^2+p^2)^{-2}=
A_d\kappa^{-\epsilon}/\epsilon$, we end up with
\begin{eqnarray}
\label{Zs'}
& & Z_{s'}=1+\left[ - {5u_0^2 \over 16D_0^2} - {u_0 s_0 \over 2D_0^2}
- {u_0 s_0 \tilde s_0 \over 2s_0' D_0^2} \right. \nonumber \\
& & \left. \qquad\qquad - {u_0 \tilde s_0' \over 2D_0^2} -
{u_0 s_0 \tilde s_0' \over 2s_0' D_0^2} - {s_0^2 \tilde s_0' \over
2s_0' D_0^2} \right] {A_d\kappa^{-\epsilon}\over\epsilon} \ .
\end{eqnarray}
\begin{figure}
  \centerline{\epsfxsize 8.5cm \epsfbox{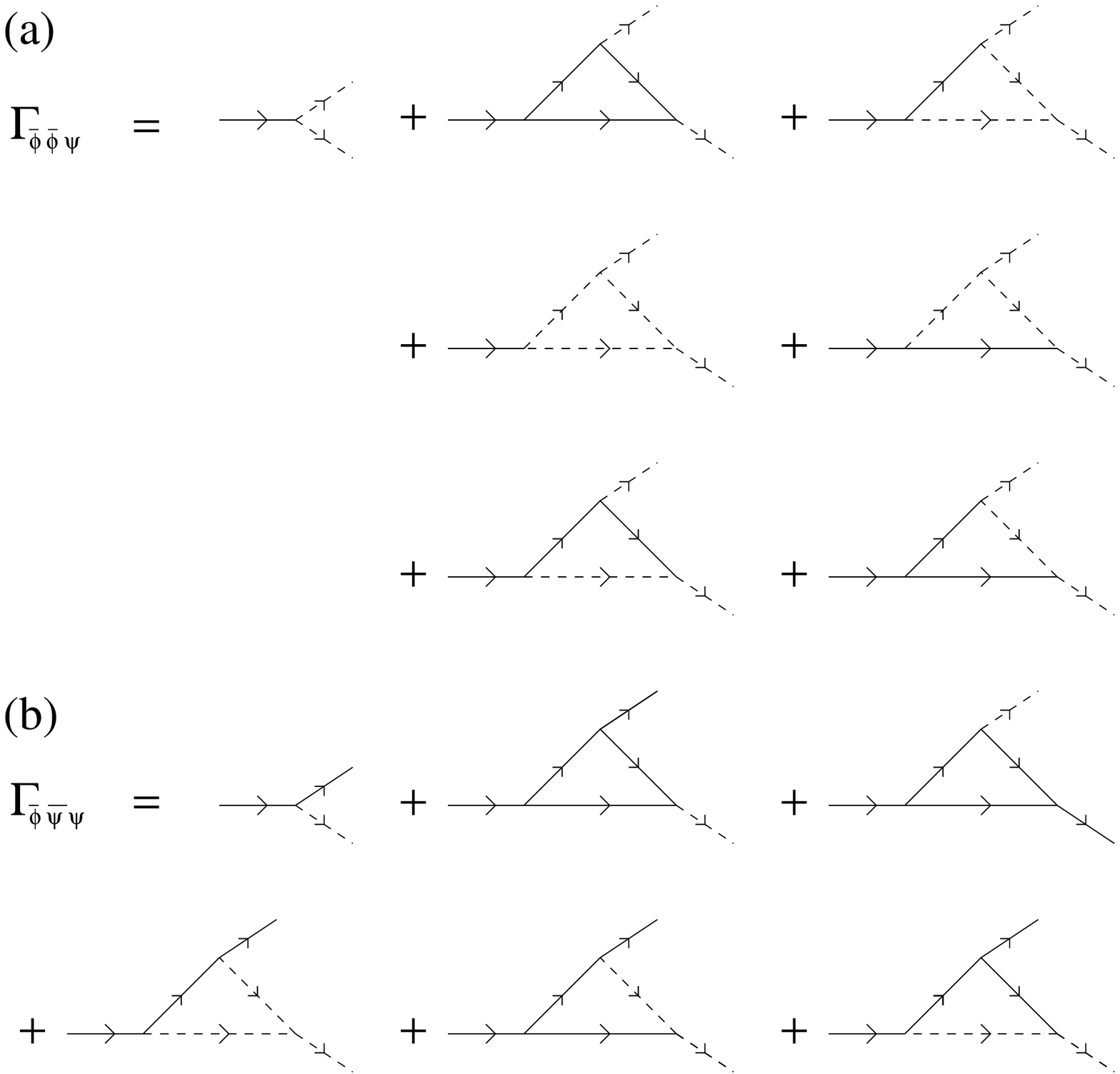}}
\vspace{3mm}
\caption{The ``mixed'' three-point vertex functions (a) 
  $\Gamma_{\bar\varphi\bar\varphi\psi}$, and (b)
  $\Gamma_{\bar\varphi\bar\psi\psi}$, to one-loop order.}
\label{genrenormdiag1}
\end{figure}
Similarly, with the aid of the diagrams shown in 
Fig.~\ref{genrenormdiag1}(b) and Fig.~\ref{genrenormdiag2}, we can 
compute the $Z$ factors for the other mixed three-point couplings,
\begin{eqnarray}
\label{Zsother}
& & Z_s = 1 + \left[ - {u_0^2 \over 16D_0^2} - {u_0 s_0 \over 4 D_0^2}
- {u_0^2 \tilde s_0 \over 4s_0 D_0^2} \right. \nonumber \\
& & \qquad\qquad\qquad\qquad \left. - {u_0 \tilde s_0' \over 2D_0^2}
\right] {A_d\kappa^{-\epsilon}\over\epsilon} \ , \\
& & Z_{\tilde s} = 1 + \left[ - {5 u_0^2\over 16D_0^2} - {u_0 s_0 
\over 2D_0^2} - {u_0 \tilde s_0'\over 2D_0^2} \right. \nonumber \\
& & \left. \quad - {u_0 \tilde s_0' s_0 \over 2 \tilde s_0 D_0^2} -
{u_0 s_0'\tilde s_0' \over 2\tilde s_0 D_0^2} - {\tilde s_0'^2 s_0
\over 2\tilde s_0 D_0^2} \right] {A_d \kappa^{-\epsilon} \over 
\epsilon} \ , \\
& & Z_{\tilde s'}=1 + \left[ - {u_0^2 \over 16D_0^2} - {u_0^2s_0' 
\over 4 \tilde s_0' D_0^2} - {u_0 s_0\over 4D_0^2} \right.\nonumber\\
& & \qquad\qquad\quad \left. - {u_0 \tilde s_0' \over 4D_0^2} \right]
{A_d\kappa^{-\epsilon}\over\epsilon} \ .
\end{eqnarray}
\begin{figure}
  \centerline{\epsfxsize 8.5cm \epsfbox{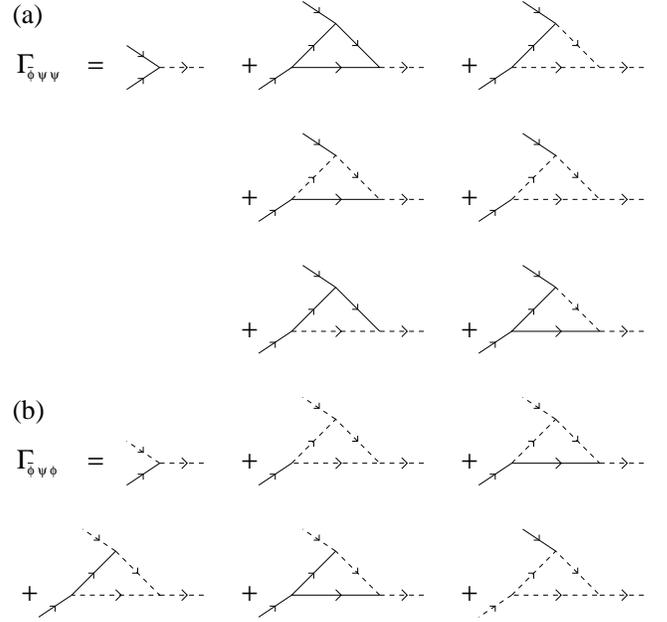}}
\vspace{3mm}
\caption{The ``mixed'' three-point vertex functions (a)
  $\Gamma_{\bar\varphi \psi\psi}$, and (b)
$\Gamma_{\bar\varphi\psi\varphi}$ to
  one-loop order.}
\label{genrenormdiag2}
\end{figure}
We note in passing that, in principle, a product of the quartic 
vertices [which we previously discarded from the action 
(\ref{action0})] and $\mu_0$ might also enter the renormalizations of 
the three-point functions.  However, we have checked that these 
additional couplings all have negative RG eigenvalues and are 
therefore irrelevant.

With the definitions $g=s/D$, $g'=s'/D$, $\tilde g=\tilde s/D$ and
$\tilde g'=\tilde s'/D$, it is now straightforward to compute the RG 
$\beta$ functions for these new variables.  Since we have
$\beta_g=\kappa\partial_{\kappa}g=g(\zeta_s-\zeta_D)$ etc., where
$\zeta_s=\kappa\partial_{\kappa}\ln(s/s_0)$ etc., we find to one-loop
order
\begin{eqnarray}
\label{betag}
& & \beta_g=-{\epsilon\over 3}(g-\tilde g)+{\sqrt{\epsilon\over3}}g
\left( {1\over 2}g+\tilde g'\right) \ , \\
\label{betag'}
& & \beta_{g'}={\sqrt{\epsilon\over 3}}g(g'+\tilde
g)+{\sqrt{\epsilon\over 3}}
\tilde g'(g+g')+{1\over 2}g^2\tilde g' \ , \\
\label{betatildeg}
& & \beta_{\tilde g}={\sqrt{\epsilon\over 3}}g(\tilde g+\tilde g') 
+{\sqrt{\epsilon\over 3}}\tilde g'(g'+\tilde g)+{1\over 2}\tilde g'^2g 
\ , \\
\label{betatildeg'}
& & \beta_{\tilde g'}={\epsilon\over 3}(g'-\tilde g')+
{\sqrt{\epsilon\over 3}} \tilde g'\left(g+{1\over 2}\tilde g'\right) 
\ ,
\end{eqnarray}
where we have already inserted the one-loop fixed point value for
$v^*=[(u/4D)^*]^2$. Notice the symmetry of the above RG $\beta$ 
functions with respect to exchanging $g \leftrightarrow \tilde g'$
and $g' \leftrightarrow \tilde g$.  We now search for fixed-point
solutions of the above equations where $\beta_g^* = \beta_{g'}^* =
\beta_{\tilde g}^* = \beta_{\tilde g'}^* = 0$.  Using 
Eqs.~(\ref{betag}) and (\ref{betatildeg'}) to eliminate $\tilde g$ 
and $g'$ from the remaining two $\beta$ functions, the above system 
of equations can be solved exactly.  After some tedious algebra, we
find two {\em fixed-line} solutions, the first one being
\begin{eqnarray}
\label{fixedlineunstable}
& & g^*=-\tilde g'^*, \quad \tilde g^*-g'^*=2g^* \ , \nonumber \\
& & {g^*}^2=2{\sqrt{\epsilon\over 3}}(g^*+g'^*) \ .
\end{eqnarray}
Computing the eigenvalues of the stability matrix
\begin{equation}
\label{stabilitymatrix}
{\partial\beta_{\gamma}\over\partial\delta} \ , \quad {\rm with} 
\quad \{\gamma\},\{\delta\}=\{g,g',\tilde g,\tilde g'\} \ , 
\end{equation}
yields the eigenvalues $0$, $0$, $-\epsilon/3$, $-\epsilon/3$.  
Hence, this first fixed line, which includes the Gaussian fixed 
points for the mixed three-point couplings, is {\em unstable}.  
The second fixed line is given by
\begin{eqnarray}
\label{fixedlinestable}
& & g'^*=\tilde g^*, \quad 
g^*+\tilde g'^*=2{\sqrt{\epsilon\over 3}} \ , \nonumber \\
& & {g^*}^2=2{\sqrt{\epsilon\over 3}}(g^*+g'^*) \ ,
\end{eqnarray}
with stability matrix eigenvalues $0$, $\epsilon/3$, $\epsilon/3$,
$4\epsilon/3$. Hence this second fixed line is {\em stable}.  Notice 
also that {\em both} of the above fixed lines satisfy the condition
\begin{equation}
\label{cancelcondition}
{\sqrt{\epsilon\over 3}}(g'^*+\tilde g^*)=-g^*\tilde g'^* \ ,
\end{equation}
which ensures the cancellation of the strongly singular (UV-divergent in
$d=2$) diagrams for the renormalization of $\mu_0$ in 
Eq.~(\ref{murenormstep1}). 

We can now insert the values of $g^*+\tilde g'^*$ at the two fixed lines
(\ref{fixedlineunstable}) and (\ref{fixedlinestable}), respectively,
into Eq.~(\ref{zetamu}), and thus obtain
\begin{equation}
\label{zetamu*}
\zeta_{\mu}^*=\cases{-2+\epsilon/6+O(\epsilon^2) & (stable line) \ , \cr
-2-\epsilon/6+O(\epsilon^2) & (unstable line) \ . }
\end{equation}
We are now in a position to {\em derive} the scaling form
(\ref{genscal}) for
the $B$ species density, postulated in the previous section. We begin by
writing down the renormalization group equation for the renormalized
field
$\langle\varphi_R\rangle$, where the angular brackets denote averaging
with
respect to the full action $S_{\rm mc}$
\begin{eqnarray}
\label{RGequation}
& &
\left(\kappa{\partial\over\partial\kappa}+\zeta_{\tau}\tau{\partial\over
\partial\tau}+\zeta_D D{\partial\over\partial D}+\zeta_v v{\partial\over
\partial v}+ \right. \nonumber \\
& & \left. +\zeta_g g{\partial\over\partial g}+\zeta_{g'}g'
{\partial\over\partial g'}+\zeta_{\tilde g} \tilde g
{\partial\over\partial\tilde g}+\zeta_{\tilde g'} \tilde g'
{\partial\over\partial\tilde g'} \right. \\
& & \left. +\zeta_{\mu}\mu{\partial\over\partial\mu}-{1\over 2} 
\zeta_{\varphi}\right)\langle\varphi_R(\kappa,\tau,D,v,\{g\},\mu,x,t)
\rangle=0 \nonumber \ ,
\end{eqnarray}
and where we have used the notation $\{g\}=\{g,g',\tilde g,\tilde g'\}$.
Defining the dimensionless field $\hat\varphi$ as
\begin{eqnarray}
\label{dimlessvarphi}
& & \langle\varphi_R(\kappa,\tau,D,v,\{g\},\mu,x,t)\rangle=\kappa^{d/2} 
\nonumber \\
& & \qquad\qquad\quad \times\hat\varphi(\tau,v,\{g\},\mu/D,\kappa x,
\kappa^2 Dt) \ ,
\end{eqnarray}
the solution of Eq.~(\ref{RGequation}) when the couplings $v,\{g\}$
have run to their fixed-point/line values is
\begin{eqnarray}
\label{RGsolution}
& & \langle\varphi_R(\kappa,\tau,D,v,\{g\},\mu,x,t)\rangle=\kappa^{d/2}
\ell^{(d-\zeta_{\varphi}^*)/2} \nonumber \\ & & \quad\times 
\hat\varphi\left(\tau \ell^{\zeta_{\tau}^*},v^*,\{g^*\},(\mu/D)
\ell^{\zeta_{\mu}^*-\zeta_D^*}, \right. \nonumber \\
& & \qquad\qquad\quad \left. 
\kappa x \ell, \kappa^2 Dt \ell^{2+\zeta_D^*} \right) \ .
\end{eqnarray}
Inserting the matching condition $\ell=|\tau|^{-1/\zeta_{\tau}^*}$, and
dropping the $v,\{g\}$ couplings, we obtain
\begin{eqnarray}
\label{scalingvarphi0}
& & \langle\varphi_R(\kappa,\tau,D,\mu,x,t)\rangle\propto
|\tau|^{-(d-\zeta_{\varphi}^*)/2\zeta_{\tau}^*} \\
& & \qquad\times\hat\varphi\left( (\mu/D)
|\tau|^{-(\zeta_{\mu}^*-\zeta_D^*)/\zeta_{\tau}^*}, 
\kappa x |\tau|^{-1/\zeta_{\tau}^*}, \right. \nonumber \\
& & \qquad\qquad\qquad\qquad \left. \kappa^2 Dt 
|\tau|^{-(2+\zeta_D^*)/\zeta_{\tau}^*} \right) \ . \nonumber
\end{eqnarray}
Identifying $\beta_1=\beta=-(d-\zeta_{\varphi}^*)/2\zeta_{\tau}^*$,
$\nu_{\perp}=-1/\zeta_{\tau}^*$, $\nu_{\parallel}=
-(2+\zeta_D^*)/\zeta_{\tau}^*$ and defining the crossover exponent as 
$\phi=(\zeta_{\mu}^*-\zeta_D^*)/\zeta_{\tau}^*$, we have
\begin{equation}
\label{scalingvarphi}
\langle\varphi_R(\kappa,\tau,D,\mu,x,t)\rangle\propto |\tau|^{\beta_1}
\hat\varphi\left({\mu/D\over|\tau|^{\phi}},
{\kappa x \over|\tau|^{-\nu_{\perp}}}, {\kappa^2 Dt \over
|\tau|^{-\nu_{\parallel}}}\right) \ ,
\end{equation}
in agreement with the scaling hypothesis (\ref{genscal}) postulated in
the previous section. Using our earlier results for the $\zeta^*$ 
functions, we find
\begin{equation}
\label{phivalues}
\phi=\cases{1+O(\epsilon^2) & (stable line) \ , \cr
1+\epsilon/6+O(\epsilon^2)  & (unstable line) \ .}
\end{equation}
Notice the absence of $O(\epsilon)$ contributions to $\phi$ at the
stable fixed line, which is due to remarkable cancellations.  This of 
course also implies that there are no logarithmic corrections to the 
crossover exponent $\phi$ in $d_c = 4$ dimensions.
 
The final step in this calculation is now to compute the exponent
$\beta_2$.  Unfortunately, in order to do this, we must first 
understand the behavior of the scaling function (\ref{scalingvarphi}) 
in the active phase.  As we shall see, it contributes non-trivial 
corrections to the exponent $\beta_2$ at $O(\epsilon)$.  Hence it is 
{\it not} sufficient simply to match the scaling function to that 
calculated in mean-field theory \cite{correc} --- such a procedure 
would miss these $O(\epsilon)$ corrections. However, before dealing
further with this active phase calculation for the $B$ species, we 
first discuss the simpler problem of pure DP in the active phase, 
which also applies to the first level (the $A$ particles) of the 
coupled DP problem.

\subsection{DP field theory: Active phase calculation}
\label{DPactivephase}


In this section, we will review the one-loop calculation of the
expectation value of the field in the active phase for the case of 
a single field, i.e., the case of (decoupled) directed percolation. 
In this way we can obtain an expression for the critical exponent 
$\beta_1 = \beta$ \cite{ABSS}.

We start with the action for a single field (see Sec.~\ref{prelim})
\begin{eqnarray}
& & S_{\rm DP}=\int \! d^dx \int \! dt \, \left\{ {\bar{\psi}}_0\left[ 
\partial_t + D_0 (r_0-\nabla ^2)\right] \psi _0 - \right. \nonumber \\ 
& & \left. \qquad\qquad\qquad\qquad\qquad 
-{\frac{u_0}2}\left( {\bar{\psi}}_0^2\psi_0-{\bar{\psi}}_0\psi_0^2
\right) \right\} \ .
\label{L1}
\end{eqnarray}
In the active phase ($r_0<0$) the expectation value of the field
$\psi_0$ is non-zero, and we define a shifted field $\psi_{c0}$ by
\begin{eqnarray}
\psi _0=v_0\ +\ \psi _{c0} \ , 
\label{shift}
\end{eqnarray}
to obtain the new action
\begin{eqnarray}
& & S'_{\rm DP}= \int \! d^dx \int \! dt \,
\left\{\bar{\psi}_0(\partial_t
\psi _{c0}-D_0\nabla ^2\psi_{c0}+D_0r_0\psi _{c0})+ \right.\nonumber \\
& & \left.\quad +\bar\psi_0(D_0r_0v_0+\frac 12u_0v_0^2) -\frac 12u_0v_0
\bar\psi_0^2+u_0v_0\bar\psi_0\psi_{c0}+ \right.\nonumber \\
& & \left.\qquad\qquad\qquad\qquad +\frac 12u_0(\psi _{c0}-\bar\psi_0)
\bar\psi_0\psi _{c0}\right\} \ .
\label{shiftedlag}
\end{eqnarray}
We now fix $v_0$ by equating the coefficient of $\bar\psi_0$ to zero.
Thus
\begin{equation}
v_0=-\frac{2D_0r_0}{u_0}=\frac{2D_0|r_0|}{u_0} \ ,  \label{v0}
\end{equation}
where $v_0$ is the classical (mean-field) value of the expectation value
of the field. Substituting these values into Eq.~(\ref{shiftedlag}), we 
obtain
\begin{eqnarray}
& & S'_{\rm DP}=\int \! d^dx \int \! dt \, \left[\bar{\psi}_0(\partial_t
\psi _{c0}-D_0\nabla ^2\psi_{c0}+D_0|r_0|\psi _{c0})+ \right. \nonumber
\\ & & \qquad\qquad +\left. \frac 12u_0(\psi _{c0}-\bar\psi_0)
\bar\psi_0\psi_{c0}-D_0|r_0|\bar\psi_0^2 \right] \ . 
\label{nlag}
\end{eqnarray}
In the following we will define $\tau _0\equiv -r_0$.  From 
Eq.~(\ref{shift}) it follows that
\begin{equation}
\langle \psi _0\rangle = v_0\ +\ \langle \psi _{c0}\rangle \ . 
\label{expval}
\end{equation}
There is only one diagram contributing to the expectation value 
$\langle \psi_{c0}\rangle $ to one-loop order, and it is depicted in 
Fig.~\ref{figact1}.
\begin{figure}
  \centerline{\epsfysize 4.5cm \epsfbox{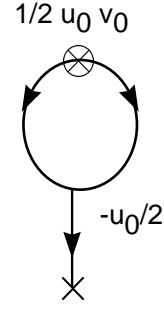}} \vspace{3mm}
\caption{One-loop diagram for $\langle \psi_{c0}\rangle$ in the pure DP
  active phase calculation.}
\label{figact1}
\end{figure}
We thus find, using dimensional regularization,
\begin{eqnarray}
\langle \psi _0\rangle & = & \frac{2D_0\tau _0}{u_0}-\frac12
\frac{u_0}{D_0} \int_p\frac 1{p^2+\tau _0} \nonumber \\
& = & \frac{2D_0\tau _0}{u_0}-\frac{u_0}{D_0}\frac{%
\Gamma (1-d/2)}{(8\pi )^{d/2}}(2\tau _0)^{d/2-1} \ .  \label{1loopb}
\end{eqnarray}
Using the relations between the bare and renormalized quantities given
in Eqs.~(\ref{rftrenorm})--(\ref{Zfactoru}), the last expression 
yields
\begin{eqnarray}
& & \langle \psi _R\rangle =\frac{2D\tau }uA_d^{1/2}\kappa^{d/2}Z_\psi
^{1/2}Z_uZ_\tau ^{-1}Z_D^{-1} \nonumber \\
& & \qquad\qquad\qquad\qquad \times\left( 1+\frac{u^2}{4\epsilon D^2}
\frac{\tau^{\epsilon /2}}{1-\epsilon /2}\right) \ .
\end{eqnarray}
We see that all the poles in $\epsilon$ cancel out as they should.  At
the fixed point $(u/D)^{*}=2(\epsilon /3)^{1/2}$, we finally find to 
leading order in $\epsilon $,
\begin{equation}
\langle \psi _R\rangle =2\left( \frac Du\right)^{*}A_d^{1/2}
\kappa^{d/2} (1+\epsilon /6)\ \tau \ \left( 1-\frac \epsilon 6
\ln \tau \right) \ .
\end{equation}
Upon exponentiating the logarithm, assuming that this can be done
unambiguously, we see that
\begin{equation}
\langle \psi_R\rangle =n_A \sim \tau ^{1-\epsilon /6} \ ,
\end{equation}
which implies $\beta_1 = \beta = 1-\epsilon /6+O(\epsilon ^2)$, as
already cited in Eq.~(\ref{DPbeta}), where the scaling relation 
$\beta = \nu_\perp (d + z - 2 + \eta_\perp) / 2$ was employed.

\subsection{Coupled DP field theory: Active phase}
\label{coupleDPactive}


We now proceed to discuss the active phase calculation for two coupled
fields, as represented by the action $S_{\rm mc}=S_{\rm{eff}}+\Delta S$ 
given in Eqs.~(\ref{action1}) and (\ref{deltaS}).  We will again restrict 
our discussion to the case where $r_A=r_B=r_0$.  Based on the mean-field 
analysis of Sec.~\ref{meanf}, we anticipate that the expectation values 
of the fields are different from zero when $r_0<0$.  Hence we define the 
shifted fields
\begin{eqnarray}
\psi _0 &=&v_{A0}+\psi _{c0} \ , \\
\bar{\psi }_0 &=&\bar{\psi }_{c0} \ , \\
\varphi _0 &=&v_{B0}+\varphi _{c0} \ ,  \label{shiftedf} \\
\bar{\varphi }_0 &=&\bar{\varphi }_{c0} \ .
\end{eqnarray}
The new action expressed in terms of the shifted fields will now involve
terms linear in $\bar{\psi }_{c0}$ and $\bar{\varphi }_{c0}$.  Equating 
the coefficients of these terms to zero fixes the constants $v_{A0}$ and
$v_{B0}$ to be the classical (mean-field) values. These are determined 
by the equations
\begin{eqnarray}
& & D_0r_0v_{A0}+\frac 12u_0v_{A0}^2 =0 \ , \label{expectpf} \\
& & D_0r_0v_{B0}+\frac 12u_0 v_{B0}^2-\mu _0v_{A0}+\frac 12
\tilde s_0v_{A0}^2+\tilde s_0'v_{A0}v_{B0} =0 \ . \nonumber
\end{eqnarray}
The solutions of these equations are
\begin{eqnarray}
v_{A0} &=&\frac{2D_0|r_0|}{u_0} \ , \\
v_{B0} &=&\sqrt{\frac{4\mu _0D_0|r_0|}{u_0^2}}+O(|r_0|) \ .
\label{approxshift}
\end{eqnarray}
The fields now have new masses given by
\begin{eqnarray}
D_0r_{A0} &=&D_0r_0+u_0v_{A0}=D_0|r_0| \ , \\
D_0r_{B0} &=&D_0r_0+u_0v_{B0}+\tilde s_0'v_{A0} \\
&\simeq &\sqrt{4\mu _0D_0|r_0|}+O(|r_0|) \ .
\label{rab}
\end{eqnarray}
\begin{figure}
  \centerline{\epsfxsize 8.5cm \epsfbox{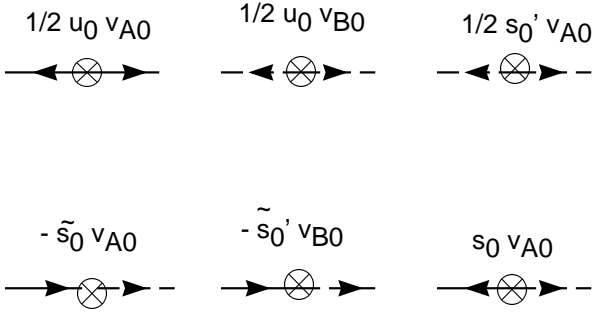}} \vspace{3mm}
\caption{New two-point vertices in the active phase coupled DP
  calculation.}
\label{figaact2}
\end{figure}
There are also various new two-point vertices as depicted in
Fig.~\ref{figaact2}, whose corresponding terms in the action are given
by
\begin{eqnarray}
&&\int \! d^dx \int dt \left[-\frac 12u_0v_{A0}\bar{\psi}_{c0}^2-
\frac12(u_0v_{B0}+s_0'v_{A0})\bar{\varphi }_{c0}^2- \right. \nonumber\\
&& \left. -(\mu _0-\tilde s_0v_{A0}-\tilde s_0'v_{B0})
\bar{\varphi }_{c0}\psi _{c0}-s_0v_{A0}\bar{\varphi }_{c0}
\bar{\psi }_{c0} \right] \ .  \label{2pt}
\end{eqnarray}
We can now proceed with the calculation of the expectation value of the
second field $\varphi_0$, by using Eq.~(\ref{shiftedf}) and computing 
$\langle \varphi_{c0}\rangle $ to one-loop order.  The full 
calculation is unnecessarily complicated, and universality dictates 
that we can calculate the critical behavior for {\em any} point on the 
stable fixed line.  However, we note that the unstable and stable fixed 
lines do not necessarily have to give the same critical behavior, and 
indeed in our case we will find that they do not.

For the stable fixed line, we have actually performed the calculation in
two different ways.  Both methods yield identical results, and this 
provides us with an important extra check on our methods.  In the first 
approach we have chosen a subspace of initial parameters
\begin{equation}
s_0=\tilde s_0',\ s_0'=\tilde s_0 \ ,  \label{mcp}
\end{equation}
with the stable fixed point $(u/D)^{*}=2\sqrt{\epsilon /3},\
g^{*}=\tilde g'^*=-2g'^*=-2\tilde g^*=\sqrt{\epsilon /3}$.  In the second 
approach we have chosen
\begin{equation}
s_0'=\tilde s_0=\tilde s_0'=0 \ ,  \label{s0}
\end{equation}
and with the stable fixed point $(u/D)^{*}=2\sqrt{\epsilon/3},\ 
g^{*}=(s/D)^{*}=2\sqrt{\epsilon /3},\ g'^*=\tilde g ^{*}=\tilde g'^*=0$.

Note that both the parameter subspaces given by Eqs.~(\ref{mcp}) and
(\ref{s0}) are closed under renormalization flows, as is evident from
a close inspection of the RG $\beta$ functions given by
Eqs.~(\ref{betag}), (\ref{betag'}), (\ref{betatildeg}) and
(\ref{betatildeg'}).  The first choice is the more natural one for
investigating the multicritical point, since when Eq.~(\ref{mcp}) is
satisfied the original action $S_{\rm mc}=S_{\rm eff}+ \Delta S$ is
invariant under a generalization of the usual DP `rapidity reversal',
namely
\begin{eqnarray}
\psi_0(x,t) &\longleftrightarrow &-\bar{\varphi}_0(x,-t) \ , \\
\bar{\psi}_0(x,t) &\longleftrightarrow &-\varphi_0(x,-t) \ ,
\label{symmetry}
\end{eqnarray}
and under renormalization this symmetry is preserved. However, the
calculation using the condition (\ref{s0}) is somewhat simpler. In
both approaches the unstable (non-trivial) fixed point is given by
$(u/D)^{*}=2\sqrt{\epsilon /3}$, and $g^{*}=g'^*=\tilde g^*=
\tilde g'^*=0$.

The diagrams contributing to the expectation value of $\langle
\varphi_{c0}\rangle $ to one-loop order are depicted in 
Fig.~\ref{figact2}.
\begin{figure}
  \centerline{\epsfxsize 8.5cm \epsfbox{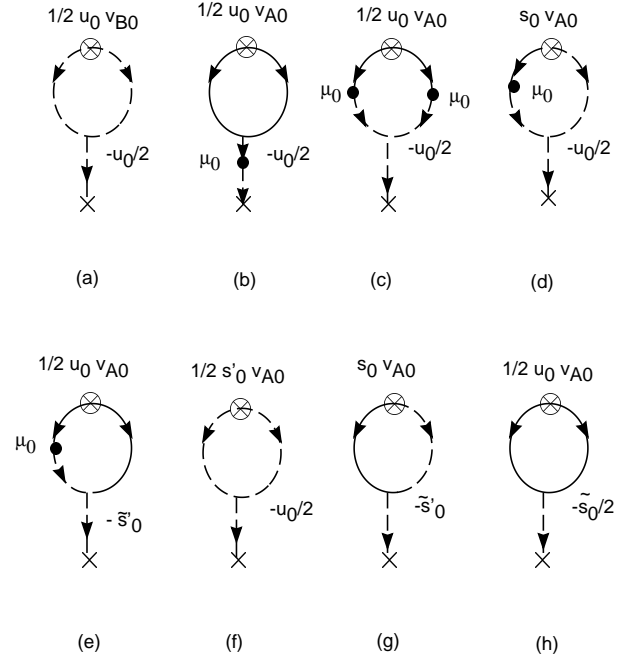}} \vspace{3mm}
\caption{
\label{ActivePhaseDiagrams}
One-loop diagrams for the computation of $\langle\varphi_{c0} \rangle$
in the coupled DP active phase calculation.}
\label{figact2}
\end{figure}
We will outline first the calculation using the second approach
described above.  In that case diagrams (e) through (h) vanish and 
one has to consider only the first four diagrams.  We thus find to 
one-loop order, after implementing Eq.~(\ref{s0}),
\begin{eqnarray}
& & \langle \varphi _0\rangle  =v_{B0}-\frac{u_0^2v_{B0}}{2D_0r_{B0}}%
I_1(r_{B0})-\frac{u_0^2v_{A0}\mu _0}{2D_0^2r_{A0}r_{B0}}I_1(r_{A0})
\nonumber \\
&&-\frac{u_0^2v_{A0}\mu _0^2}{2D_0^{}r_{B0}}I_2(r_{A0,}\ r_{B0})-\frac{%
u_0^{}s_0v_{A0}\mu _0}{D_0^{}r_{B0}}I_3(r_{B0};\ r_{A0},r_{B0})\nonumber
\\& & \qquad\qquad\qquad\qquad\qquad\qquad\qquad\qquad\qquad +\cdots \ . 
\label{expf}
\end{eqnarray}
With the use of dimensional regularization, the various integrals
appearing in Eq.~(\ref{expf}) are given by the following expressions
\begin{eqnarray}
& & I_1(a) = \frac 1{2D_0}\int_p\frac 1{p^2+a}=-\frac{A_d}{D_0\epsilon
(2-\epsilon )}\ a^{1-\epsilon /2},  \label{I1} \\
& & I_2(a,b) =\frac 1{4D_0^3}\int_p\frac 1{(p^2+a)(p^2+b)
(p^2+\frac{a+b}2)} \label{I2} \\
& & \qquad\quad\; =-\frac{A_d}{D_0^3\epsilon (2-\epsilon )}\frac 
1{(a-b)^2} \left[ a^{1-\epsilon /2}+b^{1-\epsilon /2}- \right. 
\nonumber \\
& & \left. \qquad\qquad\qquad\qquad\qquad -2\left( \frac{a+b}2 
\right)^{1-\epsilon/2}\right] , \nonumber \\
& & I_3(b;a,b) =\frac 1{4D_0^2}\int_p\frac 1{(p^2+b)(p^2+\frac{a+b}2)}
\label{I3} \\
& & \qquad\qquad\: =\frac{A_d}{2D_0^2\epsilon (2-\epsilon )}
\left[ \frac{a+b}{a-b}\left( \frac{a+b}2\right) ^{-\epsilon /2}+ \right. 
\nonumber \\ 
& & \left. \qquad\qquad\qquad\qquad\qquad\qquad +\frac{2b}{b-a}
b^{-\epsilon /2}\right] \nonumber \ .
\end{eqnarray}
We are interested in obtaining the $\epsilon$ expansions of the above
expressions in the regime where $b\gg a$ (since $r_{B0}\gg r_{A0}$ as
$r_0\rightarrow 0$).  In that case we find
\begin{eqnarray}
I_1(a) &\approx & -\frac{A_d \kappa^{-\epsilon}}{2D_0\epsilon} \left( 1+
\frac \epsilon 2\right) a\left(1-\frac \epsilon 2\ln \frac{a}{\kappa^2}
\right) +\cdots \ , \nonumber \\
I_2(a,b) &\approx &\frac{A_d \kappa^{-\epsilon}}{4D_0^3}\frac{\ln
2}b+\cdots \ ,  \label{I2s} \\
I_3(b;a,b) &\approx &\frac{A_d \kappa^{-\epsilon}}{4D_0^2\epsilon}
\left( 1+ \frac \epsilon 2-\frac \epsilon 2\ln 2\right) \left( 1-\frac 
\epsilon 2 \ln \frac{b}{\kappa^2} \right) + \nonumber \\ & & + \cdots 
\ . \nonumber
\end{eqnarray}
Let us postpone for the time being a full discussion of the term
containing $I_2$, i.e. the contribution from diagram (c) in 
Fig.~\ref{figact2}.  Notice that this term is ultraviolet finite, i.e. 
it does not involve a pole in $\epsilon$.  However, after using 
expression (\ref{I2s}) it appears to contribute a (negative) constant 
to the expectation value of $\varphi$.  This seems problematic and a 
full discussion of this point will be given in the next section.

We now define again
\begin{equation}
\tau _0=-(r_0-r_{0c}) \ ,  
\label{t0}
\end{equation}
where to first order in standard perturbation theory we have
\begin{equation}
r_{0c}=-\frac{u_0^2}{4D_0^2}\int_p\frac 1{p^2} \ .  
\label{r0cp}
\end{equation}
This integral vanishes in dimensional regularization, and hence in the
following calculation we will ignore $r_{0c}$, as was done in the pure
DP active phase computation.  When using the relations between the
bare and renormalized quantities in Eqs.~(\ref{rftrenorm}) and
(\ref{murdef}), we find to leading order in an expansion in $\tau $:
\begin{eqnarray}
\langle \varphi _R\rangle  &\approx &\frac{2\sqrt{\mu D\tau }}%
uA_d^{1/2}\kappa ^{d/2}Z_\varphi ^{1/2}Z_uZ_\mu^{-1/2}Z_\tau^{-1/2}
Z_D^{-1/2} \nonumber \\
&&\hspace{-8mm}\times \left[1+\frac{u^2}{4D^2\epsilon }(1+\epsilon /2)
\left( 1-\frac \epsilon 4\ln \frac{4\mu \tau }D\right) + \right.
\nonumber \\
&&\hspace{-8mm}\left.+\frac{u^2}{8D^2\epsilon }(1+\epsilon /2)\left(
1-\frac \epsilon 2\ln \tau\right) - \right.  \label{frexp} \\
&&\hspace{-8mm}\left. -\frac{us}{8D^2\epsilon }(1+\epsilon /2-\epsilon \ 
\ln 2/2)\left( 1-\frac\epsilon 4\ln \frac{4\mu \tau }D\right) +\cdots
\right] \ .  \nonumber 
\end{eqnarray}
A check reveals that all the poles in $\epsilon$ cancel, as they
should.  At the fixed point given by $(u/D)^{*}=(s/D)^{*}=
2\sqrt{\epsilon /3}$, we find
\begin{eqnarray}
\langle \varphi _R\rangle  &\approx &2\left( \frac Du\right) ^{*}\left(
1+\frac \epsilon 6\right) \sqrt{\frac{\mu \tau }D}  \label{frlog} \\
&&\hspace{-10mm} \times \left( 1-\frac \epsilon {12}\ln (\mu \tau /D)-
\frac \epsilon{12}\ln \tau +\frac \epsilon {24}\ln (\mu \tau /D)+\ldots
\right) \ . \nonumber
\end{eqnarray}
Assuming that only one such scaling term becomes generated, we may now
exponentiate the logarithms, and see that
\begin{eqnarray}
\langle \varphi_R \rangle &\sim &\tau ^{1/2-\epsilon /8}
(\mu /D)^{1/2-\epsilon /24}  \nonumber \\
&=&\tau^{1-\epsilon /6}\left( \tau^{-1}\mu /D\right)^{1/2-\epsilon /24} 
\ , \label{nbf}
\end{eqnarray}
from which we infer
\begin{eqnarray}
\beta _2 &=&\frac 12-\frac \epsilon 8+O(\epsilon ^2) \ , \label{b2} \\
\phi  &=&1+O(\epsilon ^2) \ ,  \label{crossover}
\end{eqnarray}
where $\phi $ is the crossover exponent, a result consistent with that
derived previously in the inactive phase.  We also see that in the 
definition
\begin{equation}
\langle \varphi _R\rangle =\tau ^{\beta _1}\widehat{\varphi }
(\tau^{-\phi }\mu /D),  
\label{scaling1}
\end{equation}
the scaling function behaves for large argument as
\begin{equation}
\widehat{\varphi }(x)\sim x^{1/2-\epsilon /24},  
\label{scalinglarge}
\end{equation}
a result that differs from the mean-field behavior at $O(\epsilon)$.  At
the upper critical dimension $d_c = 4$, a comparison with 
Eq.~(\ref{oplog}) for the first hierarchy level suggests the logarithmic 
correction 
\begin{equation}
\langle \varphi_R \rangle \sim \tau^{1/2} (- \ln \tau)^{1/4} \ .
\label{cpdplog}
\end{equation}

Let us also discuss briefly the calculation using the approach given by
Eq.~(\ref{mcp}) above.  In that case we have an additional four 
diagrams (e)-(h) to consider.  However, we will see that diagrams 
(f)-(h) contribute to higher-order terms in an expansion in $\tau $, 
beyond the leading behavior described above.  Diagram (e) gives an 
equal contribution to diagram (d), but since the value of $s$ at the 
corresponding fixed point is half of what it was in the previous 
calculation, the final result is exactly the same.  To verify these 
claims we observe that diagrams (e) to (h) contribute the following
additional terms to the r.h.s. of Eq.~(\ref{expf}),
\begin{eqnarray}
&&-\frac{u_0s_0v_{A0}\mu
_0}{2D_0r_{B0}}I_3(r_{A0};r_{B0},r_{A0})-\frac{%
u_0s_0^{^{\prime }}v_{A0}}{2D_0r_{B0}}I_1(r_{B0}) - \nonumber \\
&&-\frac{s_0^2v_{A0}}{D_0r_{B0}}I_4(r_{A0},r_{B0})-
\frac{u_0s_0^{^{\prime}} v_{A0}}{2D_0r_{B0}}I_1(r_{A0}) \ .  
\label{addterms}
\end{eqnarray}
The only new integral $I_4$ is given by
\begin{equation}
I_4(a,b)=I_1((a+b)/2) \ .  \label{I4}
\end{equation}
It is now easy to verify that the last three terms give a contribution
of $O(\tau _0)$ and higher which can be neglected in comparison with 
the $O(\sqrt{\tau_0})$ terms which we have kept. The $I_3$ contribution 
in the first term is given by
\begin{eqnarray}
I_3(a;b,a) &=&\frac{A_d}{2D_0^2\epsilon (2-\epsilon )}\left[ 
\frac{b+a}{b-a} \left( \frac{a+b}2 \right)^{-\epsilon /2} \right. 
\nonumber \\
& & \left. \qquad\qquad\qquad\qquad +\frac{2a}{a-b} a^{-\epsilon/2}
\right] . \nonumber \\
& & \hspace{-15mm} \approx \frac{A_d}{4D_0^2\epsilon }\left( 
1+\frac \epsilon 2+\frac \epsilon 2\ln 2\right) \left( 1-\frac 
\epsilon 2\ln b\right) +\cdots \ , \label{I3aab}
\end{eqnarray}
and thus diagram (e) yields essentially the same contribution as diagram
(d).  Since the value of $(s/D)^{*}$ is now one-half of its value in the
previous approach, our final result follows.

Finally, let us discuss the behavior at the unstable, non-trivial fixed
point given by $(u/D)^{*}=2\sqrt{\epsilon /3}$, and $g^{*}=g'^*=\tilde
g^*=\tilde g'^*=0$. Returning to Eq.~(\ref{frexp}), with the last term
set equal to zero, we have
\begin{equation}
\langle \varphi _R\rangle \propto \sqrt{\frac{\mu \tau }D}
\left( 1-\frac \epsilon {12}\ln (\mu \tau /D)-\frac \epsilon {12}\ln
\tau +\ldots \right) \ , 
\label{frlog1}
\end{equation}
to leading order.  Exponentiating the logarithms (again with the
assumption mentioned above) we find
\begin{eqnarray}
\langle \varphi _R\rangle  &\sim &\tau ^{1/2-\epsilon /6}(\mu /D)^{ %
1/2-\epsilon /12 }\\     \nonumber \\
&=&\tau ^{1-\epsilon /6}\left( \tau ^{-1-\epsilon /6}\mu /D 
\right)^{1/2-\epsilon /12} \ ,
\end{eqnarray}
and thus in this case $\beta _2=1/2-\epsilon /6+O(\epsilon ^2),\ \phi
=1+\epsilon /6+ O(\epsilon ^2)$ and the scaling function $\hat{\varphi}
(x)\sim x^{1/2-\epsilon /12+O(\epsilon^2)}$ for large values of the 
argument.

\subsection{Technical difficulties}
\label{techdiff}


We now return to diagram (c) of Fig.~9. Substituting the expression for 
$I_2(a,b)$ from Eq.~(\ref{I2s}) into the corresponding expression in 
Eq.~(\ref{expf}), we find that the contribution of diagram (c) to the
expectation value of $\langle \varphi _0\rangle $ is:
\begin{equation}
-\frac{u_0^2A_d\kappa ^{-\epsilon }v_{A0}}{8D_0^4r_{B0}^2}\mu _0^2\ln 2
\rightarrow -\kappa ^{d/2}A_d^{1/2}\left( \frac uD\right) ^{*}\frac{\ln
2}{16}\frac \mu D,  \label{ccont}
\end{equation}
to leading order in $\epsilon $ and in an expansion in powers of
$\sqrt{\tau}$.  This is in contrast to our expectation (backed up by our 
simulation results in Sec.~\ref{simul}) that $\langle \varphi \rangle $ 
should vanish at the transition as $\tau \rightarrow 0$.  Notice that 
this diagram is ultraviolet-finite and hence there are no poles in 
$\epsilon $.  On the other hand, the loop integral $I_2(a,b)$ is 
infrared-{\em divergent} for any $d\leq 6$, which leads to the 
behavior $I_2(a,b)\propto 1/b$ for $d=4$ when $a\rightarrow 0$. One 
can argue that for $d>4$ the prefactor of this diagram will vanish 
due to the fact that $u$ flows to the Gaussian fixed point $u^{*}=0$.  
But for $d<4$ we seem to have a problem, the origin of which is
obviously the appearance of the strongly {\em relevant} parameter $\mu$
as an effective coupling (two-point vertex) in the perturbation 
expansion.

This difficulty is somewhat reminiscent of the random-field problem,
where infrared-divergent diagrams in perturbation theory lead to a 
shift in the upper critical dimension in a $\varphi ^4$ theory from $4$ 
to $6$ \cite{AIM}.  This is due to the extra propagators in a given 
loop as compared to the pure case.  One might perhaps argue that 
because of the unidirectionality of the interaction between the two 
fields $\psi $ and $\varphi $, the $\psi $ field acts as a
spatially and temporally correlated random field from the point of view
of the $\varphi $ field, since there is no backwards feedback, meaning 
that the $\varphi $ field has no influence on the $\psi $ field. However, 
in contrast to the random field case, where the random field is taken to 
be of constant variance as one approaches the transition temperature, the
expectation value of the $\psi $ field vanishes as one approaches the
multicritical point, and this seems to soften the effect of the infrared
problem.  Another marked difference is that $\psi$ is not a quenched 
random variable in the traditional sense, since it is displaying strong
temporal fluctuations in the multicritical regime.  Since the above 
infrared-divergent integral becomes tamed at $d=6$, one might think that, 
similar to the random-field problem, one might control its divergence by 
a dimensionality shift.  However, we have not been able to construct a 
sensible field theory for this problem by choosing $d=6$ as the upper 
critical dimension.  Actually other infrared-divergent diagrams can be 
found at higher loop orders, which are even more divergent than the 
diagram just mentioned, as in, e.g., Fig.~\ref{FigIR}(a).  IR-divergent 
diagrams also appear in the expansion of other vertex functions, like 
the $\varphi \overline{\varphi }^2$ vertex to one-loop order, where one 
can easily construct diagrams with one or two insertions of the $\psi$ 
field into the $\varphi$-triangular diagram, see Fig.~\ref{FigIR}(b),(c).

\begin{figure}
  \epsfxsize=85mm \centerline{\epsffile{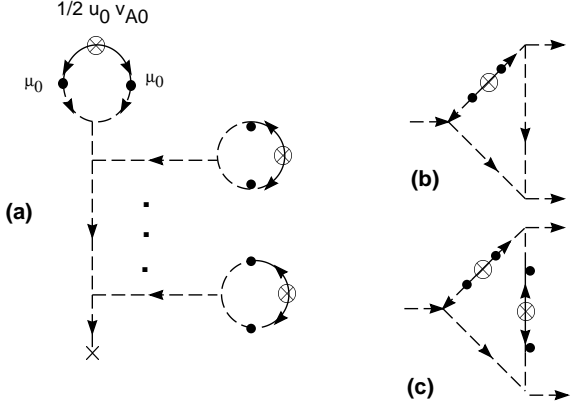}} \vspace{2mm}
\caption{
\label{FigIR}
(a) An IR-divergent diagrams at higher loop order, contributing to the 
expectation value of $\varphi$.  (b) IR-divergent diagram contributing 
to the three-point vertex function with one insertion of ${\bar \psi}^2$.  
(c) IR-divergent diagram contributing to the three-point vertex function 
with two insertions of ${\bar \psi}^2$. }
\end{figure}
Another place were a similar diagram appears is in the ``mixed'' vertex
function $\Gamma _{\overline{\varphi }\psi }$, see the last diagram of
Fig.~4 and Eq.~(\ref{murenormstep1}). If we evaluate this vertex function 
at a general ``temperature'' $\tau$, rather than at $\tau =1$ as in
Eq.~(\ref{murenormstep1}), we find that its contribution in $d=4$ 
diverges like $1/\tau$ as $\tau \rightarrow 0$.  From the active phase 
side, the same diagram diverges like $1/\sqrt{\mu \tau }$. This prevents 
one from defining the renormalized $\mu $ parameter at criticality as 
the value of this function for $\tau =0$, at least not at $q=\omega=0$. 
One can attempt to absorb this divergence by a finite (non-ultraviolet 
divergent) renormalization addition to $Z_\mu $, which will then become 
temperature dependent. Thus one would define $Z_\mu $ such that the 
renormalized $\Gamma _{\overline{\varphi }\psi }$ evaluated at
$q=\omega =0$ would be finite as $\tau \rightarrow 0$. If this is done
in the active phase we have verified that it cancels the contribution of 
the problematic contribution to $\langle \varphi _0\rangle $ against the
corresponding term in $Z_\mu$.  However, this procedure appears somewhat
artificial and not entirely satisfactory since it requires a
temperature-dependent renormalization constant. Also, to establish such
a renormalization scheme, one would have to check that this procedure
works properly to higher loop orders as well, where even more 
IR-divergent contributions are present. 

In the absence of a truly satisfactory resolution of the infrared
difficulty, one might argue that its effect is to render the $\epsilon$ 
expansion derived in the previous sections invalid.  Yet, this 
conclusion is too far reaching, since even if a well-defined field 
theory that describes the asymptotic critical behavior of all hierarchy 
levels is problematic (and may even not exist), our results are still 
likely to be valid for a range of the parameter $\tau$ close to the 
transition point, as will become clear from the following reasoning. 

We notice that the problematic IR-divergent diagrams are proportional to
higher powers of the inter-species coupling $\mu $. For example, the 
IR-problematic diagram mentioned at the beginning of this section gives 
an overall contribution proportional to $\mu $, whereas the other 
diagrams contributing to $\langle \varphi \rangle$ to one loop order 
(which were mentioned at the end of the previous section) are 
proportional to $\sqrt{\mu }$ or $\sqrt{\mu }\ln \mu $.  For small 
values of $\mu $ these problematic diagrams are therefore suppressed.
Now $\mu $ is a relevant coupling in the RG sense and thus the effective
(running) coupling increases as one goes deeper into the critical
region.  But by starting with an initially small value of $\mu$, one can 
increase the size of the region in which $\mu$ remains small.  In this 
intermediate region the IR-divergent diagrams can still be neglected and 
the scaling results obtained in the previous sections are valid.  Thus, 
our theory predicts that for small $\mu $, as $\tau $ is decreased, one 
can observe a scaling regime where the critical behavior is 
characterized by the universal exponents calculated in the previous 
sections.  In particular the exponent $\beta _2$ should be observable.
Ultimately, as $\tau$ becomes very small the field theory may break
down.  However, we cannot exclude the emergence of another non-trivial  
asymptotic scaling regime deep in the critical region. The ensuing scaling behavior might possibly be extracted by isolating the structure of the leading IR divergences to all orders and then by a resummation of the ill-defined perturbative expansion in $\mu$. However although this possibility exists, 
it cannot be substantiated at this point by more 
rigorous arguments.  What is more likely to happen is that a 
{\em non-universal} crossover behavior ensues (non-universal as a 
consequence of the breakdown of the $\epsilon$ expansion and hence the 
RG construction), which eventually terminates in asymptotically
decoupled DP behavior.  In other words, the $B$ species are no longer 
slaved to the $A$ particles, but rather behave independently such that 
their density vanishes with a power $\beta _1$.  This is supported by 
the simulation results, as will be discussed in a later section. 
Probably, at sufficiently large times, the discrete and finite number 
of $A$ particles already vanishes in the interior of comparatively 
large domains.  Consequently the $B$ particles, which were actually 
generated previously through the reaction $A \to B$, interact and
annihilate as if they were independent and decoupled from the $A$
species.  Further discussions of this issue in the light of the 
simulation results will follow at the end of the paper.

\subsection{Diagonalized theory}
\label{diagth}


We now return to another of the approaches to the coupled DP problem
mentioned in Sec.~\ref{prelim}, namely that of diagonalizing the action
(\ref{action1}) to remove the quadratic cross-term linking the $\psi_0$ 
and $\varphi_0$ fields.  If we apply the transformations
\begin{eqnarray}
\label{diagtransform}
& & \Phi_0=\varphi_0+{\mu_0\over D_0(r_A-r_B)}\psi_0 \ , \nonumber \\
& & \bar\Psi_0=\bar\psi_0-{\mu_0\over D_0(r_A-r_B)}\bar\varphi_0 \ ,
\end{eqnarray}
then the action (\ref{action1}) is transformed to
\begin{eqnarray}
\label{diagaction}
& & S_{\rm diag} = \int \! d^dx \int \! dt \left\{
{\bar \Psi}_0 \left[ \partial_t + D_0 (r_A - \nabla^2) \right] \psi_0 - 
\nonumber \right. \\
& & \left. \qquad\qquad\qquad\qquad\quad 
- {u_0 \over 2} \left[ {\bar \Psi}_0^2 \psi_0 
- {\bar \Psi}_0 \psi_0^2 \right] + \right. \\
& & \left. + {\bar \varphi}_0 \left[ \partial_t + D_0 (r_B - \nabla^2) 
\right] \Phi_0 - {u_0 \over 2} \left[ {\bar \varphi}_0^2 \Phi_0 
- {\bar \varphi}_0 \Phi_0^2 \right] - \right. \nonumber \\
& & \left. -S_0\bar\varphi_0\bar\Psi_0\psi_0 +\tilde S'_0 
\bar\varphi_0\Phi_0 \psi_0+{\tilde S_0\over 2}\bar\varphi_0\psi_0^2
-{S_0'\over 2}\bar\varphi_0^2 \psi_0\right\} \nonumber \ .
\end{eqnarray}
Here we have defined
\begin{eqnarray}
\label{newdiagvertices}
& & S_0={u_0\mu_0\over D_0(r_A-r_B)} \ , \nonumber \\
& & \tilde S_0'=-{u_0 \mu_0\over D_0(r_A-r_B)} \ , \nonumber \\
& & {\tilde S_0\over 2}={\mu_0\over 2D_0(r_A-r_B)}
\left(u_0+{u_0 \mu_0\over D_0(r_A-r_B)}\right) \ , \\
& & {S_0'\over 2}={-\mu_0\over 2D_0(r_A-r_B)}\left(u_0-{u_0\mu_0\over
D_0 (r_A-r_B)}\right) \nonumber \ .
\end{eqnarray}
Notice that this new action (\ref{diagaction}) has precisely the form of
our original full action $S_{\rm mc}$, except that the quadratic 
cross-term which linked the $\psi_0$ and $\varphi_0$ fields has been 
eliminated.  This renders the diagonalized model a special case of the
very general quadratically coupled DP processes studied by Janssen
\cite{mulcdp}.  In addition, we remark that this diagonalizing 
transformation breaks down when $r_A=r_B$, consistent with our earlier 
identification of novel multicritical behavior for $r_A=r_B=r\to 0$.

Since the action (\ref{diagaction}) is very similar to $S_{\rm mc}$, we
can quickly determine its ensuing scaling behavior.  The mixed 
three-point vertices have exactly the same fixed-line structure as 
derived in Sec.~\ref{coupleDPinactivephase}.  Thus the diagrams which 
would have generated a quadratic cross-term under renormalization (i.e., 
the $d=2$ UV-divergent diagrams in Fig.~{\ref{murenormdiag}) cancel out 
at {\em both} of the fixed lines (at least to one-loop order).  Hence, 
to this order, the DP parts of the action for the $\psi$ and $\Phi$ 
fields are entirely unaffected by the presence of the mixed three-point 
vertices, generated by the transformation (\ref{diagtransform}).  In 
that case we expect pure DP behavior on the transition lines away from 
the multicritical point, as we had earlier anticipated, and as shown
on very general grounds in Ref.~\cite{mulcdp}.

\subsection{Crossover theory}
\label{crossth}


Outside the multicritical regime, the mean-field approximation suggested 
that we should expect ordinary DP critical behavior for the $B$ species, 
if either $\tau_A > 0$ and $\tau_B \to 0$, or $\tau_B > 0$ and 
$\tau_A \to 0$, see Fig.~\ref{phdiag}.  If one starts near the 
multicritical point where both $\tau_A \to 0$ and $\tau_B \to 0$, then at 
first the multicritical density exponent $\beta_2$ should be observed, 
eventually crossing over to the ordinary DP exponent $\beta_1$, and 
similarly for the exponents $\alpha_i$, etc.  In the first crossover 
region $\tau_A > 0$ and $\tau_B \to 0$, indicated by the dashed arrow 
$B$ in Fig.~\ref{phdiag}, this scenario is rather obvious on the
mean-field level, valid for $d > d_c$: Here, the density $n_A = 0$, and
therefore the coupling between the different hierarchy levels vanishes. 
Our study of the diagonalized theory in the preceding Sec.~\ref{diagth}
established the DP character of the ensuing active/absorbing transition 
at $\tau_B = 0$ even below $d_c$, when fluctuations become dominant.

The more interesting case is the situation for $\tau_B > 0$ and $\tau_A
\to 0$, where there exists a DP phase transition for the $A$ particles. 
For $\mu_{AB} > 0$, we argued above that the $B$ species becomes 
``slaved'' to the critical $A$ species, and is driven to a 
non-equilibrium phase transition itself as the $A$ control parameter 
$\tau_A \to 0$.  In order to describe the ensuing crossover scenario, 
we apply the generalized minimal subtraction scheme of 
Ref.~\cite{crosov}, where we retain the parameter $\tau_B$ in the RG 
flow equations, and use $\tau_A = 1$ as the normalization point. The $Z$
factors then become functions of the non-linear couplings {\em and} 
$\tau_B$.  In order to simplify the calculation, we once again use the 
reduced parameter space where $g'=\tilde g=\tilde g'=0$.  However, now
we must distinguish between the non-linear DP coupling $v$ for the $A$
species, and the one for the $B$ particles, which we denote with $v'$.
The remaining  important Wilson RG functions then become
\begin{eqnarray}
& & \zeta_{\tau_A} = - 2 + 3 v \ ,
\label{crozta} \\
& & \zeta_{\tau_B} = - 2 + {3 v' \over (1 + \tau_B)^{1+\epsilon/2}} \ ,
\label{croztb} \\
& & \zeta_\mu - \zeta_D = 
- 2 - v + {\sqrt{v'} g \over (1 + \tau_B)^{1+\epsilon/2}} \ ,
\label{crozmu}
\end{eqnarray}
and the RG $\beta$ functions for the couplings $v$, $v'$, and $g$ read
\begin{eqnarray}
& & \beta_v = v \left( - \epsilon + 12 v \right) \ ,
\label{crobtv} \\
& & \beta_{v'} = v' \left( 
- \epsilon + {12 v' \over (1 + \tau_B)^{1+\epsilon/2}} \right) \ ,
\label{crobvp} \\
& & \beta_g = g \left( - {\epsilon \over 2} 
+ 2 v + {\sqrt{v'} g \over (1 + \tau_B)^{1+\epsilon/2}} \right) \ .
\label{crobtg}
\end{eqnarray}
Of course, the $A$ species DP fixed point remains $v^* = \epsilon / 12$
with $\zeta_{\tau_A}^* = - 2 + \epsilon / 4$.  However, the 
{\em effective} $B$ coupling 
$v'(\ell) / [1 + \tau_B(\ell)]^{1+\epsilon/2} \to 0$ asymptotically as
$\ell \to 0$ (since $v'(\ell) \sim \ell^{-\epsilon}$) and hence
$\tau_B(\ell) \sim \ell^{-2}$ according to Eqs.~(\ref{crobvp}) and 
(\ref{croztb}).  Thus $\beta_g \to - (\epsilon / 3) g$, and 
$g(\ell) \to 0$ as well.  Consequently,
$\zeta_{\mu}^* - \zeta_D^* = - 2 - \epsilon / 12$, and we recover
\begin{equation}
\phi = (\zeta_\mu^* - \zeta_D^*) / \zeta_{\tau_A}^* = 1 + \epsilon / 6 
+ O(\epsilon^2) \ ,
\label{croscr}
\end{equation}
i.e., the result given in Eq.~(\ref{phivalues}) at the unstable fixed
line.  Notice that here the critical exponents for the $B$ species are 
given by $\nu_\perp = - 1 / \zeta_{\tau_A}^*$ etc., and thus take on 
the usual DP values (\ref{DPeta}) to (\ref{DPbeta}).

In the active phase, we may confirm this by direct calculation, as in
Secs.~\ref{DPactivephase} and \ref{coupleDPactive}.  From the diagrams 
in Fig.~\ref{figact2}, for finite $\tau_B$, and in essentially the 
reduced parameter space, only graph (b) survives. After inserting the 
$p_c$ shift, and multiplying by the appropriate renormalization 
constants (in the generalized minimal subtraction scheme), the equation 
of state in terms of the renormalized quantities assumes the form
\begin{eqnarray}
&&\langle \varphi_R \rangle \tau_B |\tau_A| \left[ 1 - {4 v \over
\epsilon} \left( 1 + {\epsilon \over 2} - {\epsilon \over 2} 
\ln |\tau_A| \right) \right] \sim \nonumber \\
&&\qquad\qquad \sim 2 \sqrt{v} {\mu \over D} \langle \psi_R \rangle^2\ . 
\label{croeqs}
\end{eqnarray}
At the DP fixed point $v^* = \epsilon / 12$, we exponentiate the
logarithm on the left-hand side to $|\tau_A|^{1 - \epsilon/6} = 
\tau_A^{\beta_1}$, and use $\langle \psi_R \rangle \sim 
\tau_A^{\beta_1}$, which leads us to 
\begin{equation}
\langle \varphi_R \rangle \sim 2 \sqrt{v^*} \, {\mu \over D} \,
{\tau_A^{\beta_1} \over \tau_B} \ ,
\label{crobt1}
\end{equation}
which obviously generalizes the mean-field result (\ref{denDPB}).

\subsection{Higher hierarchy levels}
\label{hihierar}


We end this section on applications of the renormalization group to
coupled DP with some remarks on the behavior of higher hierarchy 
levels $i > 2$.  As pointed out already in Sec.~\ref{meanf}, while 
generically the transitions from the active to the absorbing inactive 
phase of particle species $i$ are of the DP universality class, with 
the critical density exponent $\beta_1$, one finds the two-level 
hierarchy exponent $\beta_2$ whenever the first two critical points 
coincide, $r_A = r_B$.  However, further special multicritical
behavior appears at higher levels if additional $r_k$ become equal.  
In mean-field theory, one has $\beta_i = 1/2^{i-1}$ for 
$r_1 = \ldots = r_i$.  In the two-level hierarchy $(i=2)$ 
multicritical regime, one finds a strong {\em downward} 
renormalization of $\beta_1$ and $\beta_2$ due to fluctuations, see
Eqs.~(\ref{DPbeta}) and (\ref{b2}).

In order to see what happens at higher hierarchy levels, let us briefly
consider the three-species coupled DP process in the vicinity of the
multicritical regime, and for small transmutation rates.  Notice that 
in this situation fluctuations will not only generate the vertices 
corresponding to Eq.~(\ref{deltaS}) on each adjacent level, but also the
indirect transmutation $A \to C$, as well as all possible three-point 
vertices {\em unidirectionally} coupling the levels (and in principle 
additional higher order non-linearities, which however turn out to be 
irrelevant).  In order to simplify the analysis, we merely use the 
reduced parameter space analogous to keeping only the new vertex $s_0$ 
in the two-level process.  This leaves us with the following effective
action (with the fields $\bar \phi_0$ and $\phi_0$ describing the
particle species $C$),
\begin{eqnarray}
\label{3levact}
S_{\rm eff} & & = \int \! d^dx \int \! dt \, \Bigl\{ {\bar \psi}_0 
\left[ \partial_t + D_0 (r_A - \nabla^2) \right] \psi_0 - \nonumber \\
& & \qquad\qquad - {u_0 \over 2} \left( {\bar \psi}_0^2 \psi_0 
- {\bar \psi}_0 \psi_0^2 \right) + \\
& & + {\bar \varphi}_0 \left[ \partial_t + D_0 (r_B - \nabla^2) \right] 
\varphi_0 - {u_0 \over 2} \left( {\bar \varphi}_0^2 \varphi_0 
- {\bar \varphi}_0 \varphi_0^2 \right) - \nonumber \\
& &\qquad \qquad - \mu_0 {\bar \varphi}_0 \psi_0 - s_0 {\bar \varphi}_0 
{\bar \psi}_0 \psi_0 + \nonumber \\ 
& & + {\bar \phi}_0 \left[ \partial_t + D_0 (r_C - \nabla^2) \right] 
\phi_0 - {u_0 \over 2} \left( {\bar \phi}_0^2 \phi_0 
- {\bar \phi}_0 \phi_0^2 \right) - \nonumber \\
& &\qquad \qquad - \mu_0' {\bar \phi}_0 \varphi_0 - t_0 {\bar \phi}_0 
{\bar \varphi}_0 \varphi_0 - \nonumber \\
& &\qquad - {\bar \mu}_0 {\bar \phi}_0 \psi_0 - {\bar t}_0 {\bar \phi}_0 
{\bar \psi}_0 \psi_0 - \rho_0 {\bar \phi}_0 {\bar \varphi}_0 \psi_0
\Bigr\} \ .
\nonumber
\end{eqnarray}

Clearly, if we just consider the $B / C$ reactions, then a fully
analogous calculation as in Sec.~\ref{coupleDPinactivephase} yields
\begin{equation}
\label{bctrren}
\zeta_{\mu'} - \zeta_D = - 2 - v + \sqrt{v} \, h \ ,
\end{equation}
where $h = t/D$.  Hence with $v^* = \epsilon/12$, the crossover exponent 
at the associated two-level multicritical point is
\begin{equation}
\label{crphip}
\phi' = (\zeta_{\mu'}^* - \zeta_D^*)/\zeta_\tau^* = 1 + O(\epsilon^2)\ ,
\end{equation}
computed at the stable fixed point $h^* = 2 \sqrt{\epsilon/3}$ (which is
the remnant of the stable fixed line in the reduced parameter space).  We 
may wonder now if the unidirectional, sequential coupling of three 
hierarchy levels leads to a further novel crossover exponent associated 
with the transmutation $A \to C$.  Thus, we compute the renormalization 
of the vertex function $\Gamma_{{\bar \phi} \psi}$, which leads to
\begin{equation}
\label{actrren}
\zeta_{\bar \mu} - \zeta_D = - 2 - v + \sqrt{v} \, (f + {\bar f}) \ ,
\end{equation}
where $f = {\bar t} / D$ and ${\bar f} = \rho \mu' / D {\bar \mu}$.  We
now merely need the renormalizations of both $f$ and ${\bar f}$.  In just 
the same manner as in the two-level calculation, one finds from the 
renormalization of $\Gamma_{{\bar \phi} {\bar \psi} \psi}$,
\begin{equation}
\label{betaf3}
\beta_f = f (- \epsilon/2 + 2 v + \sqrt{v} \, f) 
= f (-\epsilon/3 + \sqrt{\epsilon/12} \, f) \ ,
\end{equation}
which obviously has the IR-stable fixed point $f^* = 2 
\sqrt{\epsilon/3}$.  The only really novel renormalization concerns the 
vertex function $\Gamma_{{\bar \phi} {\bar \varphi} \psi}$, yielding
\begin{equation}
\label{abcren}
\zeta_\rho - \zeta_D = - \epsilon/2 - 2 v + \sqrt{v} \, (g + h + f) \ .
\end{equation}
After using $\zeta_{\mu'}-\zeta_{\bar \mu} = \sqrt{v} (h-f-{\bar f})$, 
this leads to
\begin{eqnarray}
\label{betabf3}
\beta_{\bar f} &=& {\bar f} [- \epsilon/2 - 2 v + \sqrt{v} \, (g + 2h 
- {\bar f})] = \nonumber \\
&=& {\bar f} (\epsilon/3 - \sqrt{\epsilon/12} \, {\bar f}) \ .
\end{eqnarray}
Comparing with Eq.~(\ref{betaf3}), we see that the non-trivial fixed
point ${\bar f}^* = 2 \sqrt{\epsilon/3}$ is {\em unstable}, whereas 
${\bar f}^* = 0$ is stable for $d < 4$.  Consequently, the three-species 
vertex $\Gamma_{{\bar\phi} {\bar \varphi} \psi}$ vanishes asymptotically, 
i.e., becomes irrelevant, which implies
\begin{equation}
\label{crphib}
{\bar \phi} = (\zeta_{\bar \mu}^* - \zeta_D^*)/\zeta_\tau^* 
= 1 + O(\epsilon^2) \ , 
\end{equation}
identical with the one-loop values of the crossover exponents $\phi$ and
$\phi'$.  Coupling to an additional hierarchy level therefore does 
{\em not} introduce a novel crossover exponent in the multicritical 
regime, at least to $O(\epsilon)$. We suspect that this is actually true 
to higher orders in $\epsilon = 4-d$ and for higher levels $i > 3$ as 
well.

On the other hand, below the critical dimension $d_c=4$ the density
exponents $\beta_i$ are affected by the fluctuation corrections to the 
scaling functions, and are not simply determined by a scaling relation 
like Eq.~(\ref{mcscai}) \cite{correc}.  In fact, were such a 
renormalization contribution from the scaling function absent, we would 
arrive at $\beta_i = 1/2^{i-1} - \epsilon/6$ to one-loop order, with the 
$O(\epsilon)$ correction {\em independent} of the level index $i$.  This 
would predict that near four dimensions, i.e., for any $0 < \epsilon 
\ll 1$, $\beta_i$ should become negative for sufficiently large $i$.
The correct result (\ref{b2}) for the second hierarchy level, however,
shows that the $O(\epsilon)$ correction is actually smaller than for the 
first level.  Presumably, on each successive level the $O(\epsilon)$ 
corrections are further reduced, such that all the density exponents 
remain positive.  A detailed computation of $\beta_3$, which requires an 
explicit study of the active phase, already becomes a rather tedious 
affair, and we leave our discussion of higher hierarchy levels with this 
speculation.

\section{Numerical results}
\label{simul}


In order to support our field-theoretical results, we study a
unidirectionally coupled hierarchy of DP models using Monte-Carlo
simulations. There is a large variety of DP lattice models that can be
used for this purpose. One of the simplest and most efficient
realizations is {\em directed bond percolation} on a tilted square
lattice \cite{dirper}. In this model, neighboring sites are connected
by directed bonds which are open with probability~$p$ and closed
otherwise. Activity percolates through open bonds along a given
direction which is usually interpreted as the direction of
time. Labelling different rows of sites by a discrete time
variable~$t$, directed bond percolation may be equivalently defined as
a stochastic cellular automaton with parallel update rules mapping the
system's configuration at time~$t$ probabilistically onto a set of new
configurations at time~$t+1$. In one spatial dimension, directed bond
percolation is just a special case of the Domany--Kinzel cellular
automaton~\cite{domkin} which is known to be one of the most efficient
realizations of DP on a computer. Another advantage of using directed
bond percolation is the availability of very precise estimates for the
critical percolation threshold $p_c(d)$ in $d \leq 2$ spatial
dimensions.  Currently the best estimates are
$p_c(1)=0.644\,700\,15(5)$ \cite{crit1d} for $d=1$ and
$p_c(2)=0.287\,338(6)$ \cite{crit2d} for $d=2$, respectively. To our
knowledge, the percolation threshold for ($3+1$)-dimensional directed
bond percolation has not been estimated before. Using standard methods
we find the value $p_c(3)=0.13235(20)$.
\begin{figure}
  \epsfxsize=85mm \centerline{\epsffile{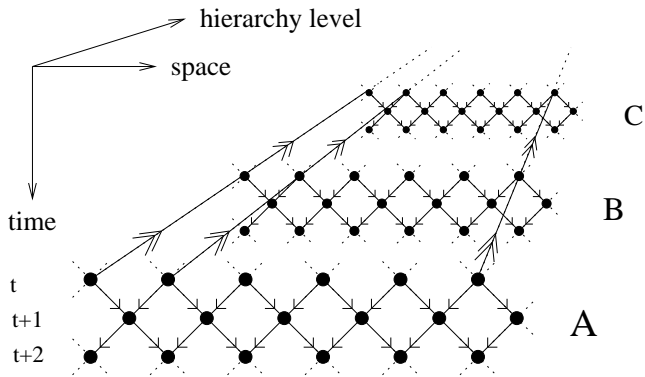}} \vspace{2mm}
\caption{
\label{FigBondDP}
Schematical illustration of a coupled hierarchy of three directed bond
percolation processes in 1+1 dimensions. Activity percolates along the
direction of time through bonds (simple arrows) which are open with
probability $p$. The three subsystems are coupled by instantaneous
transfer of activity (double arrows) to the corresponding site of the
next subsystem.}
\end{figure}

Coupled DP may be realized on a computer by simultaneously evolving
several directed bond percolation models which are coupled without
feedback according to the principles outlined in the Introduction (see
Fig.~\ref{FigBondDP}). For simplicity, we work in the limit of
infinite coupling strength $\mu=\infty$, i.e., active sites in one of
the subsystems instantaneously turn the corresponding sites of the
next subsystem into the active state.  Alternatively, we could also
apply a finite coupling strength (probabilistic transfer of activity
to the next subsystem), or explicit particle transmutation $A
\rightarrow B, B \rightarrow C, \ldots$. Although these variants are
expected to have the same critical properties, their initial crossover
times into the scaling regime are typically longer. Therefore we
restrict the numerical analysis to the case of infinite coupling
strength.

In principle it would be possible to simulate arbitrarily many
hierarchy levels. However, in order to reach the scaling regime, the
particle densities have to be low enough. It turns out that already at
the fourth level the particle density is rather high, which makes it
extremely difficult to determine critical exponents. For this reason
our numerical simulations are restricted to three hierarchy levels.

\subsection{Numerical estimation of the critical exponents}
In order to estimate the critical exponents of coupled DP, we employ
two standard numerical methods for systems with phase transitions into
absorbing states. On the one hand, we use steady-state simulations in
the active phase in order to directly determine the exponents
$\beta_k$. On the other hand, dynamical simulations~\cite{dynsim} at
the multicritical point render a set of dynamic exponents which in
turn determine the exponents $\nu_{\perp,k}$ and $\nu_{\parallel,k}$.
 
\paragraph{Steady state simulations in the active phase.}
On the multicritical line the stationary particle densities $n_k$ are
expected to scale as $n_k \sim (p-p_c)^{\beta_k}$. By measuring $n_k$
in a sufficiently large system, it is therefore possible to directly
estimate the exponents~$\beta_k$. The accuracy of the results depends
on the accessible range of $\Delta p=p-p_c$ in the simulation. In
$d=1$ spatial dimension the minimal value of $p-p_c$ is predominantly
limited by the equilibration time $T_{\rm equ}$ which has to be larger
than the temporal correlation length
$\xi_{\parallel,k}\sim(p-p_c)^{-\nu_{\parallel,k}}$, whereas in $d
\geq 2$ dimensions the main limitation is the system size $N_{\rm
max}$, which has to be larger than $\xi_{\perp,k}^d \sim
(p-p_c)^{-d\nu_{\perp,k}}$ (see Table~\ref{TableN1}). The measurements
for $d=1$ are shown in Fig.~\ref{FigSimul}a. From the slopes of the
lines averaged over one decade we estimate the exponents $\beta_i$,
whose values are listed in Table~\ref{TableN1}.
\begin{table}
\begin{tabular}{|c||c|c|c|c|}
                & $d=1$         & $d=2$         & $d=3$      &
$d=4-\epsilon$
\\\hline
$\Delta p_{\rm min}$& $0.0004$      & $0.0008$      & $0.0016$   &\\
$N_{\rm max}$       & $4000$        & $100^2$       & $35^3$     &\\
$T_{\rm equ}$       & $10^5$        & $10^4$        & $10^3$    
&\\\hline
$\beta_1$       & $0.280(5)$    & $0.57(2)$     & $0.80(4)$
&$1-\epsilon/6+O(\epsilon^2)$\\
$\beta_2$       & $0.132(15)$   & $0.32(3)$     & $0.40(3)$
&$1/2-\epsilon/8+O(\epsilon^2)$\\
$\beta_3$       & $0.045(10)$   & $0.15(3)$     & $0.17(2)$     
&$1/4-O(\epsilon)$ \\\hline
$\beta_{\rm DP}$    & $0.2765$      & $0.584$       & $0.81$       &
\end{tabular}
\caption{
\label{TableN1}
Steady-state simulation results.}
\end{table}
\begin{figure}
  \epsfxsize=85mm \centerline{\epsffile{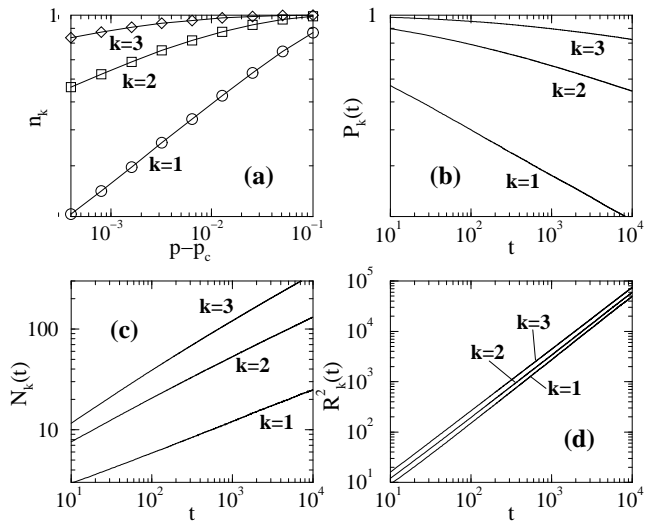}} \vspace{2mm}
\caption{
\label{FigSimul}
Monte Carlo simulations of coupled DP in $1+1$ dimensions.  a):
Steady-state simulations in the active phase.  b)-d): Dynamical
simulations at criticality.  Time is measured in Monte Carlo steps.}
\end{figure}
\paragraph{Dynamical simulations at the multicritical point.}
The most precise estimates for the critical exponents of systems with
phase transitions into absorbing states are usually obtained by
dynamical simulations~\cite{dynsim}. Starting from an initial state
with a single particle (active seed), the system evolves at the
critical point and generates a spatio-temporal cluster of active sites
whose size and lifetime is finite.  Survival probability, mass, and
mean square spreading of the cluster vary algebraically with certain
dynamical exponents which in turn are related to the exponents
$\beta$, $\nu_\perp$, and $\nu_\parallel$. In order to apply this
technique to coupled DP, we prepare an initial state with a single
$A$ particle at the origin and perform the simulation at the
multicritical point. The properties of the resulting cluster are
analyzed separately for each particle species, i.e., we measure the
survival probability $P_k(t)$, the number of $k$-particles (cluster
mass) $N_k(t)$, and the mean-square spreading from the origin
$R^2_k(t)$ averaged over all runs that survived at level $k$ up to
time $t$. At the multicritical point, these quantities are expected to
scale as
\begin{equation}
\label{DynamicalQuantities}
P_k(t) \sim t^{\delta_k} \ , \quad
N_k(t) \sim t^{\eta_k} \ , \quad
R^2_k(t) \sim t^{2/z_k} \ , 
\end{equation}
where $\delta_k=\beta_k/\nu_{\parallel,k}$ and
$z_k=\nu_{\parallel,k}/\nu_{\perp,k}$. Here, $\eta_k$ is the so-called
critical initial slip exponent~\cite{crslp1,crslp2} which will be
discussed below (not to be confused with the static correlation
function Fisher exponent $\eta$).

\begin{table}
\begin{tabular}{|c||l|l|l|l|}
                & $d=1$         & $d=2$         & $d=3$       &
$d=4-\epsilon$
\\\hline
$t_{max}$       & $10^4$        & $10^3$        & $300$ &
\\\hline
$\delta_1$      & $0.157(4)$    & $0.46(2)$     & $0.73(5)$
&$1-\epsilon/4+O(\epsilon^2)$\\
$\delta_2$      & $0.075(10)$   & $0.26(3)$     & $0.35(5)$
&$1/2-\epsilon/6+O(\epsilon^2)$\\
$\delta_3$      & $0.03(1)$     & $0.13(3)$     & $0.15(3)$   
&$1/4 - O(\epsilon)$\\\hline
$\eta_1$        & $0.312(6)$    & $0.20(2)$     & $0.10(3)$
&$\epsilon/12+O(\epsilon^2)$\\
$\eta_2$        & $0.39(2)$     & $0.39(3)$     & $0.43(5)$   
&$1/2+O(\epsilon^2)$\\
$\eta_3$        & $0.47(2)$     & $0.56(4)$     & $0.75(10)$     
&$3/4-O(\epsilon)$\\\hline
$2/z_1$         & $1.26(1)$     & $1.10(2)$     & $1.03(2)$     &\\
$2/z_2$         & $1.25(3)$     & $1.12(3)$     & $1.04(2)$
&$1+\epsilon/24+O(\epsilon^2)$\\
$2/z_3$         & $1.23(3)$     & $1.10(3)$     & $1.03(2)$     &
\end{tabular}
\caption{
\label{TableN2}
Dynamical simulation results.}
\end{table}
The temporal variation of the quantities~(\ref{DynamicalQuantities})
measured in a three-level coupled directed bond percolation model in
$1+1$ dimensions is shown in Fig.~\ref{FigSimul}b-d. Similar
simulations were performed in two and three spatial dimensions. From
the slopes of the lines averaged over two decades we estimate the
critical exponents $\delta_k$, $\eta_k$ and $z_k$, which are
summarized in Table~\ref{TableN2}. Notice that $z_1$, $z_2$, and $z_3$
assume the same values within the numerical error. Inserting the
previous estimates for $\beta_k$, we can compute the scaling exponents
$\nu_{\parallel,k}=\beta_k/\delta_k$ and
$\nu_{\perp,k}=\nu_{\parallel,k}/z_k$ separately for each level $k$ in
the hierarchy (see Table~\ref{TableN3}).  Within numerical error they
coincide with the DP exponents $\nu_\perp$, $\nu_\parallel$, and $z$,
as predicted by the field-theoretical RG calculation.
\begin{table}
\begin{tabular}{|c||l|l|l|l|}
                        & $d=1$         & $d=2$     & $d=3$ &
$d=4-\epsilon$
\\\hline
$\nu_{\perp,1}$         & $1.12(4)$     & $0.70(4)$     & $0.57(4)$  
&\\
$\nu_{\perp,2}$         & $1.11(15)$    & $0.69(15)$    &
$0.59(8)$           
&$1/2+\epsilon/16+O(\epsilon^2)$\\
$\nu_{\perp,3}$         & $0.95(25)$    & $0.65(15)$    & $0.62(9)$  
&\\\hline
$\nu_{\perp,{\rm DP}}$  & $1.0968$      & $0.734$       & $0.57$     
&\\\hline
$\nu_{\parallel,1}$     & $1.78(6)$     & $1.24(6)$     & $1.10(8)$  
&\\
$\nu_{\parallel,2}$     & $1.76(25)$    & $1.23(17)$    &
$1.14(15)$      
&$1+\epsilon/12+O(\epsilon^2)$\\
$\nu_{\parallel,3}$     & $1.50(40)$    & $1.15(30)$    & $1.21(15)$ 
&\\\hline
$\nu_{\parallel,{\rm DP}}$ & $1.7338$   & $1.295$       & $1.09$      &
\end{tabular}
\caption{
\label{TableN3}
Derived critical exponents.}
\end{table}

\paragraph{General problems.}
%
\begin{figure}
\epsfxsize=85mm \centerline{\epsffile{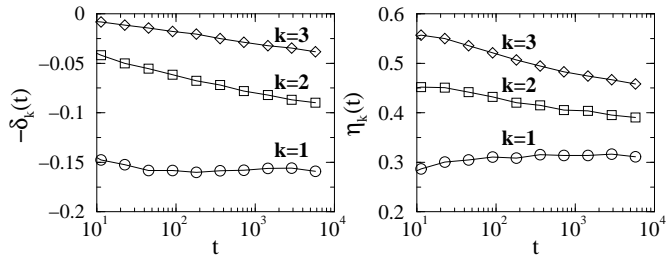}} \vspace{2mm}
\caption{
\label{FigDrift}
Local slopes of the lines in Fig.~\ref{FigSimul}b-c. The slow drift of
the slopes at higher levels indicates that the scaling regime is not
yet reached in the present simulations. (Time is measured in Monte
Carlo steps.)}
\end{figure}

The extensive simulations reveal an unexpected deviation from ideal
scaling at higher levels. As can be seen in Fig.~\ref{FigSimul}, the
lines for $k>1$ are in fact not perfectly straight but slightly
bent. In order to illustrate these deviations, we determined the local
slope of the survival probabilities
\begin{equation}
\delta_k(t) = \frac{\log_{10}P_k(2t)-\log_{10}P_k(t)}{\log_{10} 2}
\end{equation}
and similarly $\eta_k(t)$ in a ($1+1$)-dimensional system. After
sufficiently long times, these quantities should become constant and
equal to the exponents $\delta_k$ and $\eta_k$, respectively. As
expected, the lowest level reaches the scaling regime after a short
time (see Fig.~\ref{FigDrift}). At higher levels, however, there is a
considerable drift of the local slope that extends over the entire
temporal evolution. Similar drifts can be observed in all other
quantities which involve the density exponents
$\beta_2,\beta_3,\ldots$.  The deviations indicate that the scaling
regime, especially at the third level, has not yet been reached. By
estimating the critical exponents at higher levels, we therefore
encounter considerable systematical errors which may even exceed the
statistical error margins. A careful numerical analysis shows that the
drift in the local slopes is neither related to finite-size effects
nor to deviations from criticality. At present the origin of these
deviations from ideal scaling is not yet entirely clear. It might
perhaps be a signature of the IR-divergent diagram (c) of
Fig.~\ref{ActivePhaseDiagrams} (see Sect.~\ref{techdiff}). As
mentioned above, these graphs are connected with the appearance of
additional powers of the {\em relevant} coupling $\mu$, and we suspect
that this drift in the scaling exponents signals that our first-order
perturbation theory with respect to the transmutation rate becomes
ultimately insufficient, and some appropriate resummation of the
expansion in $\mu$ would in fact be required.

Furthermore, we remark that in a simulation based of course on a
finite number of particles, ultimately one would expect a crossover to
the {\em decoupled} situation, namely when the $A$ species, whose
density decays {\em faster} at the multicritical point, has already
died out.  It might well be possible that in a fluctuation-dominated
regime this effect sets in much earlier, provided there emerge large
regions which have already become depleted of the $A$ particles. Thus,
one explanation of the drifts visible in Fig.~\ref{FigDrift} could be
that this crossover region to the asymptotic decoupled regime has
already been reached. The universal exponents predicted by the field
theory would then apply only to an intermediate scaling regime. We
note that the field theory calculation, being based on a continuum
description of coarse-grained particle densities, cannot easily
account for this finiteness of the particle number.

In the case of coupled annihilation processes (see Sec. VI), where
similar deviations occur, the intermediate scaling regime can be
clearly identified in numerical simulations. In particular it is
observed that the size of the scaling regime grows as the coupling
strength decreases. We have also performed simulations of coupled DP
with reduced coupling strength (probabilistic transfer of activity to
the next level). Unfortunately, the initial crossover into the
intermediate scaling regime grows rapidly as the coupling strength is
reduced which makes it impossible to identify the boundaries of the
intermediate scaling regime.

\subsection{Critical initial slip}
When a DP process starts from random initial conditions at very low
density, the particles are initially separated by empty intervals of a
certain typical size. During the temporal evolution these particles
generate individual clusters which are initially separated. Therefore
the average particle density first {\em increases} as $n(t) \sim
t^\eta$ --- a phenomenon which is referred to as the {\em critical
initial slip} of non-equilibrium systems~\cite{crslp1}.  Later, when
the growing clusters begin to interact with each other, the DP process
crosses over to the usual decay $n(t) \sim t^{-\beta/\nu_{\parallel}}$.  
Dynamical simulations starting from a single particle represent the 
extreme case where the critical initial slip extends over the entire 
temporal evolution.

In ordinary DP, the critical initial slip exponent $\eta$ is related
to the other bulk exponents through the hyperscaling relation
$2\delta+\eta=d/z$~\cite{dynsim,crslp2}.  In the case of coupled DP we
would therefore naively expect that the critical initial slip
exponents $\eta_k$ are related to the other exponents by
$2\delta_k+\eta_k=d/z$. However, the numerical estimates in
Tables~\ref{TableN2} and \ref{TableN3} do {\em not} satisfy this
scaling relation at higher levels $k>1$. Instead they seem to fulfil
the {\em generalized hyperscaling relation} introduced in
Ref.~\cite{mendes} in the context of systems with many absorbing
states,
\begin{equation}
\label{hyperscaling}
\delta_{\rm DP} + \delta_k + \eta_k = d/z_k \ .
\end{equation}
Here $\delta_{\rm DP}$ denotes the exponent $\beta/\nu_\parallel$ of
ordinary directed percolation. In fact, inserting the estimates of
$\delta_k$ and $z_k$ for $d=1$, Eq.~(\ref{hyperscaling}) predicts the
values $\eta_1=0.314(4)$, $\eta_2=0.398(10)$, and $\eta_3=0.443(20)$,
which are in fair agreement with the estimations in
Table~\ref{TableN3}.

In fact, the above scaling relation~(\ref{hyperscaling}) may be
derived fairly simply, starting from an appropriate scaling form for
the two point correlation function. If we initiate the cluster
starting from a single localized seed, then the density of $B$
particles at a later time will have the following form:
\begin{equation}
\label{densitycorr}
n_2(x,t)\sim|\tau|^{2\beta_1}\hat f_1\left({\mu/D\over|\tau|^{\phi}},
{x\over|\tau|^{-\nu_{\perp}}},{Dt\over|\tau|^{-\nu_{\parallel}}}
\right) \ .
\end{equation}
Note that, even though this is a scaling form for the {\it density},
it has the structure of a two point correlation function. Roughly
speaking, the prefactors in Eq.~(\ref{densitycorr}) may be interpreted
as follows: one factor of $|\tau|^{\beta_1}$ comes from the
probability that the cluster is still alive at time $t$, whilst the
second factor comes from the probability that the point $(x,t)$ is a
member of that cluster. At criticality we find, by
integrating~(\ref{densitycorr}) over space
\begin{equation}
\label{intdensitycorr}
t^{\eta_2}\sim t^{(d\nu_{\perp}-2\beta_1)/\nu_{\parallel}}\hat f\left(
{\mu/D\over (Dt)^{-\phi/\nu_{\parallel}}}\right) \ ,
\end{equation}
where we have also used one of the definitions from
Eq.~(\ref{DynamicalQuantities}).  Assuming that the scaling function
$\hat f$ in the multicritical regime behaves in the same way as in
Sec.~{\ref{coupleDPactive}, we then end up with the scaling
relation~(\ref{hyperscaling}) for the second hierarchy level. The
extension to higher levels works in an exactly similar manner.

\subsection{Susceptibility to an external field}

%
\begin{figure}
  \epsfxsize=75mm \centerline{\epsffile{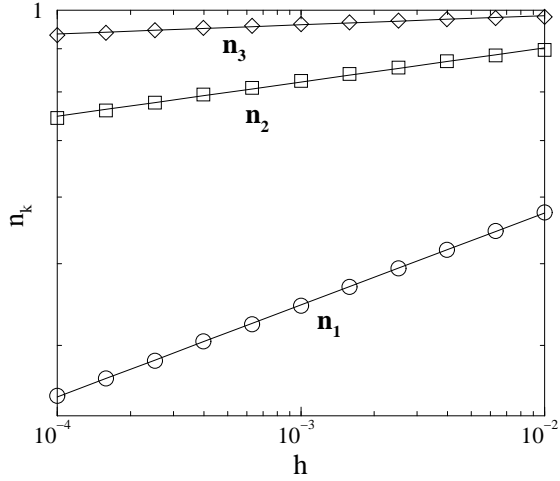}} \vspace{2mm}
\caption{
\label{FigSuscept}
Susceptibility of multicritical coupled DP to an external field. The
figure shows the stationary particle densities $n_k$ versus the rate
$h$ for spontaneous particle creation.}
\end{figure}
Coupled DP may be subjected to an external field $h>0$ by adding a
term $h\int d^dx \int dt \,\bar{a}$ to the action~(\ref{action0}). In
the particle interpretation, the field corresponds to a spontaneous
creation of $A$ particles at rate $h$ during the temporal
evolution. This means that all subsystems approach a fluctuating
steady state, irrespective of the value of $p$. We are particularly
interested in the response of $n_k$ to the external field at the
multicritical point. For ordinary DP it is known that $n_1 \sim
h^\gamma$, where $\gamma=\beta/(d \nu_\perp+\nu_{\parallel} -\beta)$
is the susceptibility exponent. For coupled DP we can derive a similar
relation, starting from the scaling form for the (steady-state)
density
\begin{equation}
\label{densfieldscaling}
n_2\sim|\tau|^{\beta_1}\hat g_1 \left(
{\mu/D\over|\tau|^{\phi}},{|\tau|\over h^{1/\Delta}}\right) \ ,
\end{equation}
where the DP exponent $\Delta$ can be shown to equal
$\Delta=d\nu_{\perp}+\nu_{\parallel}-\beta_1$.  In the limit
$|\tau|\to 0$, we thus have
\begin{equation}
\label{densfieldscaling1}
n_2\sim h^{\beta_1/\Delta}\hat g\left({\mu/D \over h^{\phi/\Delta}}
\right) \ .
\end{equation}
Hence, using the results of Sec.~\ref{coupleDPactive} and generalizing
to the $k$-th level of the hierarchy, we have
\begin{equation}
n_k \sim h^{\gamma_k} \,, \qquad
\gamma_k = \frac{\beta_k}{d \nu_\perp+\nu_{||}-\beta_1} \ .
\end{equation}
In order to verify this scaling relation, we repeat the steady-state
simulation at the multicritical point in presence of spontaneous
creation of $A$ particles. The results in $1+1$ dimension are shown in
Fig.~\ref{FigSuscept}. From the slopes of the lines we estimate the
susceptibility exponents
\begin{equation}
\gamma_1 = 0.109(2) \ , \,\,
\gamma_2 = 0.045(4) \ , \,\,
\gamma_3 = 0.014(2) \ .
\end{equation}
On the other hand, the above scaling relation yields the values
$\gamma_1 = 0.107(2)$, $\gamma_2 = 0.051(6)$, and $\gamma_3 =
0.018(5)$, which are in fair agreement with the simulation results.

\subsection{Crossover phenomena near the multicritical point}

%
\begin{figure}
  \epsfxsize=75mm \centerline{\epsffile{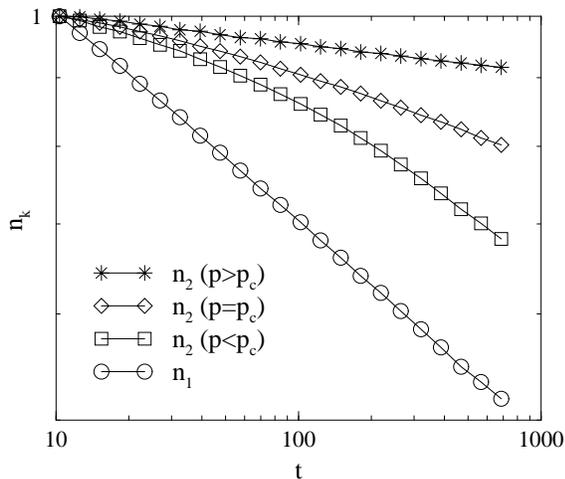}} \vspace{2mm}
\caption{
\label{FigCrossover}
Crossover effects near the multicritical point. The figure shows the
particle densities $n_k$ vs time in a two-level system starting from a
fully occupied lattice, normalized at $t=10$. Level $A$ ($k=1$) is
always critical, while level $B$ ($k=2$) is either evolving in the
active, critical, or inactive regime. Initially the decay of $B$
particles is the same in all cases. Later the system crosses over to a
different behavior where the $B$ particles become independent or
slaved to the $A$ particles, respectively. (Time is measured in Monte
Carlo steps.)}
\end{figure}
Numerical simulations near the critical point reproduce the crossover
scenario predicted by the mean-field approximation. As an example, we
consider the two crossovers along the dashed arrows $A$ and $A/B$ in
the mean-field phase diagram of Fig.~\ref{phdiag}. To this end, we
simulate a two-level system in one spatial dimension. The percolation
probability $p_1$ of level $A$ is always at the critical point,
whereas level $B$ is simulated slightly below and above criticality
$(p_2=p_c\pm 0.05)$. The numerical results are shown in
Fig.~\ref{FigCrossover}. As expected, for $p_2<p_c$ the density of $B$
particles first decays like $n_2(t) \sim t^{-\beta_2/\nu_\parallel}$
and then crosses over to a dynamical state where the $B$ particles
become ``slaved'' to the $A$ particles, such that $n_2(t) \sim n_1(t)
\sim t^{-\beta_1/\nu_\parallel}$. On the other hand, for $p_2>p_c$ the
$B$ subsystem crosses over into a state with a constant density where
the $B$ particles become independent of the $A$ particles. Thus the
crossover effects are in qualitative agreement with the mean-field and
RG predictions of Secs.~\ref{meanf} and \ref{crossth}.

\section{Applications}
\label{appl}

The most natural applications of coupled DP are to growth processes in
which the layers at different heights represent different subsystems
in the hierarchy.  The dynamical rules for adsorption and desorption
in these models have to be implemented in such a way that neighboring
layers are effectively coupled in one direction without feedback.  The
phase transition then emerges as a roughening transition from a smooth
phase to a rough phase.  The known examples include so-called
polynuclear growth models (PNG)~\cite{PNG}, a special class of
solid-on-solid (SOS) models~\cite{Alon}, and certain models for fungal
growth~\cite{Lopez}.  Another interesting realization of coupled DP is
the spreading of activity next to the ``light cone'' in stochastic
cellular automaton models with parallel update rules.

\subsection{Roughening transitions in SOS models}

Coupled DP was first identified in a particular SOS model~\cite{Alon}
which exhibits a roughening transition even in one spatial dimension.
The active phase of coupled DP corresponds to a smooth phase where the
interface is pinned to a spontaneously selected layer.  On the other
hand, the inactive phase of coupled DP corresponds to a roughening
interface which propagates at finite velocity.

The unrestricted version of the SOS model is defined on a
one-dimensional lattice of~$N$ sites, $i=1\ldots N$, with associated
height variables~$h_i$, which may take values $0,1 \ldots \infty$.
The dynamical rules are defined through the following algorithm: At
each update a site~$i$ is chosen at random.  Then an atom is adsorbed,
\begin{equation}
\label{dyn1}
h_i \rightarrow h_{i}+1 \; \; \mbox {with probability } q \ ,
\end{equation}
or desorbed from the edge of an island (plateau),
\begin{eqnarray}
\label{dyn2}
h_i \rightarrow \mbox{min}(h_{i},h_{i+1}) \; \; 
\mbox {with probability } (1-q)/2 \ ,
\\
\label{dyn3}
h_i \rightarrow \mbox{min}(h_{i},h_{i-1}) \; \; 
\mbox {with probability } (1-q)/2 \ .
\end{eqnarray}
When the growth rate~$q$ is low, the desorption processes (\ref{dyn2})
and (\ref{dyn3}) dominate.  If all the heights are initially set to
the same value $h_0$, this layer will remain the bottom layer of the
interface.  Small islands will grow on top of the bottom layer, but
will quickly be eliminated by desorption at the island edges.  Thus,
the interface is effectively anchored to its bottom layer, and the
growth velocity, defined as the rate of increase of the minimum height
of the interface, is zero in the thermodynamic limit.  As $q$ is
increased, the size of islands created on top of the lowest layer
increases. Above~$q_c$, the critical value of~$q$, the islands merge
and new layers are formed at a finite rate, giving rise to a non-zero
interface velocity in the thermodynamic limit.

The key feature of this model is that atoms may desorb only at the
{\em edges} of plateaus, i.e., at sites which have at least one
neighbor at a lower height.  In experiments this would correspond to a
system where the binding energy in completed layers is much larger
than at the edges of plateaus.  Furthermore, the dynamical processes
at a given layer are independent of the processes at higher layers.
In particular, the temporal evolution at the bottom layer is decoupled
from all other processes at higher layers.  In fact, one can show that
the dynamics of the bottom layer can be mapped onto a one-dimensional
contact process which is known to belong to the DP universality
class~\cite{Alon}.  Identifying blank sites at the bottom layer as $A$
particles, the adsorption process (\ref{dyn1}) may be interpreted as
the annihilation of $A$ particles, while the desorption process
(\ref{dyn2})--(\ref{dyn3}) corresponds to $A$ particle production.
Similarly, the dynamical processes at the following layers may be
associated with the particle species $B,C,D,\ldots$.
 
It is important to note that the state of a site in coupled DP is
characterized by presence or absence of various particle species,
while sites of a growth model are associated with a single quantity,
namely the height $h_i$.  To connect the two descriptions, we have to
assume that the coupling constant $\mu$ is infinite such that
particles at level $k$ instantaneously create particles at level
$k+1$.  In this case the state of a site in coupled DP is fully
characterized by the index of the lowest active level in the
hierarchy, which then corresponds to the height of the interface in
the growth model.  Therefore, the order parameters $n_A,n_B,n_C,\ldots
= n_1,n_2,n_3\ldots$ are defined by
\begin{equation}
\label{OrderSOS}
n_k=\frac{1}{N}\,\sum_i \, \sum_{h=h_0}^{h_0+k-1}\,\delta_{h_i,h} \ ,
\end{equation}
that is, $n_k$ is the density of sites whose heights are {\em less}
than $h_0+k$, where $h_0$ denotes the height of the bottom layer.  By
definition, the densities obey the inequality $n_{k} \leq n_{k+1}$.

The above growth model is invariant under global shifts of the heights
$h_i \rightarrow h_i+a$.  This symmetry is spontaneously broken in the
(coupled DP) active phase where the system selects a particular
reference height as the bottom layer.  In the (coupled DP) inactive
phase, however, the interface becomes rough and propagates at finite
velocity, i.e., active DP processes subsequently enter the absorbing
state.  The growth velocity $v$ is inversely proportional to the
average survival time of the lowest-lying DP process, hence $v \sim
(q-q_c)^{\nu_{||}}$.  Numerical simulations confirm that the critical
behavior of the first few layers in the growth model is indeed the
same as in coupled DP.  In particular, the exponents $\beta_k$ for the
density of sites at the first few layers are in agreement with the
numerical estimates in the present work.

Alternatively, one may study the same model with the additional
restriction $|h_i-h_{i\pm 1}|\leq 1$.  In that case the layers are no
longer coupled without feedback.  For example, if a site with height
$h_i=1$ has neighbors at heights $h_{i-1}=0$ and $h_{i+1}=2$, the atom
at site $i$ cannot desorb from the surface.  Using the language of
coupled DP, this means that the presence of $C$ particles prevents the
$A$ particles from producing offspring.  Surprisingly, the numerical
estimates of the exponents $\beta_k$ indicate that the critical
behavior of the system is still that of coupled DP.  Thus it seems
that certain realizations of ``inhibiting'' feedback from higher
levels to lower ones do not destroy the universal properties of
coupled DP.  Rather, the essential precondition for coupled DP seems
to be the existence of a hierarchy of absorbing subspaces, i.e.,
inactive levels must not be activated by higher levels.

\subsection{Polynuclear growth models}

In PNG models~\cite{PNG} a similar scenario arises, but in this case
the coupled DP behavior occurs at the highest levels of the interface.
As in the previous case, the PNG models may be defined on a square
lattice with associated height variables~$h_i$.  The key feature of
these models is the use of {\em parallel updates} which gives rise to
a maximum velocity of the interface.  One of the most popular PNG
models is defined through the following dynamical rules.  In the first
half time step atoms ``nucleate'' stochastically at the surface by
\begin{equation}
\label{nucleation}
h_i\left(t+\frac12\right) = \left\{ \begin{array}{ll}
h_i(t)+1 \ , & \mbox {with probability } p \\
h_i(t)   \ , & \mbox {with probability } 1-p \ . \\
\end{array} \right.
\end{equation}
In the second half time step the islands grow deterministically until
they coalesce
\begin{equation}
\label{coalescense}
h_i(t+1) = \max_j \left[ h_i \left( t+\frac12 \right) , h_j 
\left( t+\frac12 \right) \right] \ ,
\end{equation}
where $j$ runs over the nearest neighbors of site $i$.  Starting from
a flat interface $h_i(0)=0$, the sites at maximal height $h_i(t)=t$
may be considered as active sites of a DP process.  Obviously
Eq.~(\ref{nucleation}) turns active into inactive sites with
probability $1-p$, while offspring production is realized by the
process~(\ref{coalescense}).  Therefore, if $p$ is large enough, the
interface is smooth and propagates with velocity $1$.  Below a
critical threshold, however, the density of active sites at the
maximal height $h_i(t)=t$ vanishes, and the growth velocity is smaller
than $1$.  Identifying the sites with $h_i=t$ as $A$ particles, those
with $h_i \geq t-1$ as $B$ particles etc., the dynamical processes
resemble the rules of coupled DP.  The corresponding order parameters
are defined by
\begin{equation}
\label{OrderPNG}
n_k=\frac{1}{N}\,\sum_i \, \sum_{h=0}^{k-1}\,\delta_{h_i,t-h} \ .
\end{equation}
Thus PNG models may be interpreted as a realization of coupled DP in a
{\em co-moving} frame.  An exact mapping relating PNG models and the
previously discussed SOS models (where coupled DP resides in a fixed
frame next to the bottom layer) was proposed in Ref.~\cite{Alon}.  It
should be emphasized that the existence of a roughening transition in
PNG models requires the use of parallel updates.  If random-sequential
updates are used, the transition is lost, and the interface is always
rough since then there is no maximum velocity.

\subsection{Models for fungal growth}

Recently, L\'opez and Jensen~\cite{Lopez} introduced a class of models
for the growth of colonial organisms, such as fungi and bacteria.  The
models are motivated by recent experiments~\cite{Sams} with the yeast
{\em Pichia membranaefaciens} on solidified agarose film.  Depending
on the concentration of polluting metabolites, different front
morphologies were observed.  The aim of the models is to explain these
morphological transitions on a qualitative level.

\begin{figure}
  \epsfxsize=85mm \centerline{\epsffile{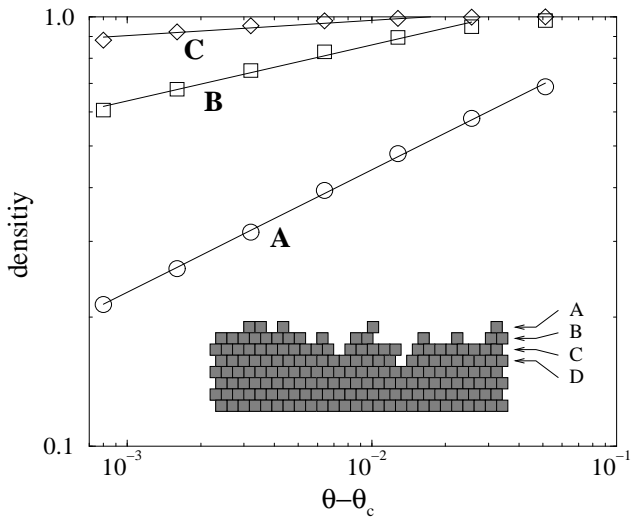}} \vspace{2mm}
\caption{
\label{FigFungus}
Coupled DP in a model for fungal growth. The graph shows the density
of the first three levels propagating at maximal velocity as a
function of $\theta-\theta_c$ in the smooth phase. Power-law fits are
used to estimate the exponents $\beta_k$ (see text). The inset shows a
typical configuration of the growing front near criticality.  }
\end{figure}

The model for fungal growth is defined on a triangular
($1+1$)-dimensional lattice whose sites are either occupied or vacant.
Growth of the colony occurs because of the division of individual
cells, i.e., only nearest neighbors of occupied sites can become
occupied. T he model evolves by {\em parallel} updates.  To mimic
realistic cells, it is assumed that cell division is less likely in
young cells.  To this end, the simulation keeps track of the age
$a_j(t)$ of occupied sites.  The probability $P_i(t)$ for a vacant
site $i$ to become occupied in the next time step depends on the total
age $A_i(t)=\sum_{\langle i,j \rangle} a_j(t)$ of the occupied nearest
neighbors of site $i$.  Using the functional dependence
$P_i(t)=\tanh[\theta A_i(t)]$, a roughening transition was observed at
$\theta_c=0.183(3)$.  Investigating clusters of sites growing at
maximal velocity, some of the critical properties at the transition
could be related to DP~\cite{Lopez}.  It was argued that this
roughening transition could be the essential mechanism behind the
morphological transitions observed in experiments.

Clearly the above fungal model and the PNG models are very similar in
character.  They both work with parallel updates and exhibit a
roughening transition which is related to DP.  Here we will present
numerical evidence to show that the fungal growth model is actually a
realization of coupled DP, in spite of complicated details such as the
age-dependent rates for cell division and interface overhangs.  The
order parameters $n_k$ may be defined as
\begin{equation}
n_k=\frac{1}{N} \, \sum_i \, N_i(t) \, \delta_{y_i,t-k+1} \ ,
\end{equation}
where $N_i(t)=0,1$ denotes the occupation number of site $i$, and
$y_i$ is the height coordinate of the $i$-th site.  We have
numerically measured the densities $n_1,n_2,$ and $n_3$ near
criticality in the smooth phase (see Fig.~\ref{FigFungus}).  Our
estimates $\beta_1=0.28(2)$, $\beta_2=0.13(2)$, $\beta_3=0.04(2)$ are
in agreement with the numerical results of section~\ref{simul}.

As in the PNG models, the existence of a roughening transition in the
model for fungal growth requires the use of parallel updates.  For
random sequential updates there is no such transition and the
interfaces are always rough.  However, random sequential updates seem
to be a more appropriate description of the experiments in
Ref.~\cite{Sams}, since realistic cells do not divide synchronously.
Therefore it is still unclear to what extent the roughening transition
of the model in Ref.~\cite{Lopez} is related to morphological
transitions in realistic fungal growth.

\subsection{Critical behavior near the light cone in spreading 
            processes with parallel dynamics}

Let us finally consider a directed bond percolation process on a
tilted square lattice in $d+1$ dimensions, which may be understood as
a stochastic cellular automaton evolving by parallel
updates~\cite{domkin}.  Starting from a single active seed such a
cellular automaton generates a cluster of active sites.  For maximal
percolation probability $p=1$ this cluster is compact and has the
shape of a pyramid.  This means that all sites within the {\em light
cone} (the surface of the pyramid) are activated.

\begin{figure}
  \epsfxsize=85mm \centerline{\epsffile{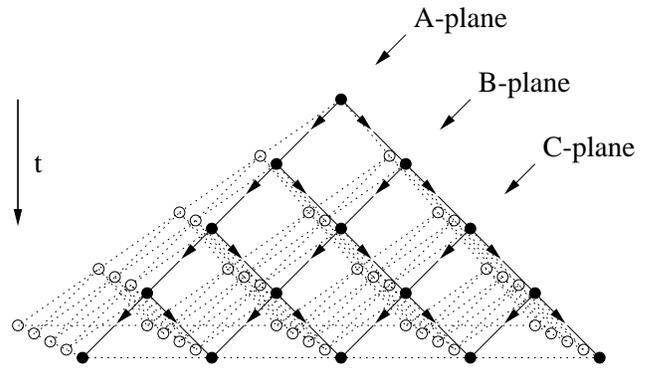}} \vspace{2mm}
\caption{
\label{FigLightCone}
Realization of a unidirectionally coupled hierarchy of
($1+1$)-dimensional DP processes in a ($2+1$)-dimensional directed
bond percolation process with parallel updates. The figure shows the
``light cone'' starting from a single site. The subsystems
$A$,$B$,$C$,$\ldots$ correspond to tilted planes as indicated by the
arrows.  }
\end{figure}

Apart from the usual phase transition, DP models with parallel updates
in $d \geq 2$ spatial dimensions exhibit a {\em second} transition,
where the clusters detach from their light cone.  In the case of
($2+1$)-dimensional directed bond percolation this transition takes
place at $p=p_s \simeq 0.6447 > p_c$.  As illustrated in
Fig.~\ref{FigLightCone}, the dynamical processes near the light cone
constitute a unidirectionally coupled hierarchy of DP processes in
$d-1$ spatial dimensions.  The lowest hierarchy level $A$ corresponds
to the sites in the light cone.  Clearly these sites are decoupled
from the interior of the pyramid.  The following hierarchy levels
$B$,$C$,$\ldots$ correspond to parallel planes as indicated in
Fig.~\ref{FigLightCone}.  Since activity can only percolate forward in
time, these planes are coupled in only one direction without feedback.
The dynamical processes within each subsystem are precisely those of a
directed bond percolation process on a tilted square lattice in $d-1$
spatial dimensions.  Therefore, the numerical value of $p_s$ in $d$
spatial dimensions coincides with the usual transition point $p_c$ in
$d-1$ dimensions.  This explains the numerical value $p_s \simeq
0.6447$.

\section{Coupled Annihilation Reactions}
\label{related}


We finally return to the question of whether new dynamic universality
classes can be constructed by the unidirectional coupling of known
non-equilibrium processes.  We have seen that in the case of {\em
linearly} coupled directed percolation, the ensuing hierarchical
structure leads to the emergence of multicritical behavior at a
special point in control parameter space, described by the novel
density exponents $\beta_i$ and $\alpha_i$.  Similarly, we expect
identical qualitative features for the closely related problem of
linearly coupled dynamic (isotropic) percolation processes (albeit
there one of the non-linear vertices is non-local in time
\cite{dynper}).  This is to be contrasted with the very general {\em
quadratically} coupled multi-color DP processes studied recently by 
Janssen, where ordinary DP critical behavior is found \cite{mulcdp}.

The simplest non-trivial case, however, would be to
consider a stochastic process which is generically scale-invariant,
i.e., where no tuning to a special critical point is required. An
example of such a system is provided by the simple diffusion-limited
two-particle annihilation reaction \cite{annihi,masfth}
\begin{equation}
\label{annrec}
A + A \to \emptyset \quad {\rm with \ rate \ } \lambda_A \ .
\end{equation}
The corresponding mean-field rate equation for the particle density $A$
reads
\begin{equation}
\label{annmfr}
{\partial n_A(x,t) \over \partial t} = 
D \nabla^2 n_A(x,t) - 2 \lambda_A n_A(x,t)^2 \ ,
\end{equation}
which is solved at long times $t \to \infty$ by $n_A(t) \sim t^{-1}$.
Power counting shows that this result is expected to be correct for
dimensions $d > d_c=2$.

In low-dimensional systems, however, fluctuations and the emerging
particle-anticorrelations become important, and the density decay
exponent is reduced.  In order to include these fluctuations
consistently, one may derive the following field theory from the
classical master equation \cite{masfth},
\begin{equation}
\label{annfth}
S = \int d^dx \int dt \Bigl[ {\hat a} ( \partial_t - D \nabla^2 ) a
- \lambda_A ( 1 - {\hat a}^2 ) a^2 \Bigr] \ , 
\end{equation}
where we have omitted boundary terms stemming from the initial
configuration, as well as terms related to the projection state (see
Ref.~\cite{masfth}).  When the action (\ref{annfth}) is expanded about
the stationary solution ${\hat a} = 1$, the classical field equation
for $a(x,t)$ yields precisely the mean-field rate equation
(\ref{annmfr}). The entire field theory (\ref{annfth}) can also be
recast in the form of a Langevin equation for the field $a(x,t)$,
although this field is related to the true density field $n_A(x,t)$ in
a rather non-trivial way \cite{jcardy,masfth,howard}.

The structure of the field theory (\ref{annfth}) is very simple, as no
diagrams can be constructed that would renormalize the free diffusion
propagator $(-i\omega + D q^2)^{-1}$.  Furthermore, the entire
perturbation series for the annihilation vertices is readily summed
via a geometric series, or through solving the ensuing Bethe-Salpeter
equation. Hence the scaling behavior of the density is known exactly.
The final result is \cite{masfth}
\begin{equation}
\label{annden}
n_A(t) \sim \left\{ \begin{array}{ll}
t^{-d/2} &{\rm for \ } d < 2 \ , \\
t^{-1} \ln t &{\rm for \ } d = d_c = 2 \ , \\
t^{-1} &{\rm for \ } d > 2 \ . \end{array} \right.
\end{equation}

Let us now consider a hierarchy of such annihilation processes,
\begin{equation}
\label{annrecB}
B + B \to \emptyset \quad {\rm with \ rate \ } \lambda_B \ ,
\end{equation}
etc., unidirectionally coupled via the branching reaction
\begin{equation}
\label{anncp}
A \to A + B \quad {\rm with \ rate \ } \sigma_{AB} \ .
\end{equation}
The choice of this specific coupling can be motivated as follows.  If
the $A$ species were not to appear on the right-hand-side of the
reaction (\ref{anncp}), then this would constitute a spontaneous death
process for the $A$ particles, immediately leading to an exponential
density decay.  However, on the lowest hierarchy level, we want to
retain all the features of the uncoupled reactions [especially the
power-law decay of Eq.~(\ref{annden})].  Also, we want to keep the
coupling reaction linear in the particle density $n_A$, as in our
earlier analysis of coupled DP.  Thus, the reaction (\ref{anncp})
feeds additional particles into level $B$, which in mean-field theory
is described by the rate equation
\begin{eqnarray}
\label{annmfrB}
{\partial n_B(x,t) \over \partial t} = &&D \nabla^2 n_B(x,t) - 2
\lambda_B n_B(x,t)^2 \nonumber \\ &&+ \sigma_{AB} n_A(x,t) \ .
\end{eqnarray}
Obviously for long times (and consequently for low densities), the $B$
particles are now slaved by the $A$ species, and their density
``adiabatically'' follows $n_A(t)$,
\begin{equation}
\label{slden}
n_B(t) \approx \left( {\sigma_{AB} \over 2 \lambda_B} n_A(t)
\right)^{1/2} \sim t^{-1/2} \ .
\end{equation}
As is to be expected, the branching process (\ref{anncp}) considerably
slows down the decay on level $B$.  Within mean-field theory, a
straightforward generalization to higher hierarchy levels leads to
$n_i(t) \sim t^{-\alpha_i}$, with $\alpha_i = 1/2^{i-1}$ on level $i$.

In order to include fluctuation effects, we write down the action
corresponding to the coupled reactions (\ref{annrec}),
(\ref{annrecB}), and (\ref{anncp}), setting 
$\lambda_A = \lambda_B = \lambda$, and $\sigma_{AB} = \sigma$:
\begin{eqnarray}
\label{cpafth}
S &&= \int d^dx \int dt \Bigl[ {\hat a} ( \partial_t - D \nabla^2 ) a
- \lambda ( 1 - {\hat a}^2 ) a^2 + \\ &&+ {\hat b} ( \partial_t - D
\nabla^2 ) b - \lambda ( 1 - {\hat b}^2 ) b^2 + \sigma (1 - {\hat b}) 
{\hat a} a \Bigr] \ . \nonumber
\end{eqnarray}
Notice that the $A$ propagator has now acquired a formal mass term 
$\sigma$, as opposed to the $B$ particles, for which we still the have 
the massless diffusion propagator (again, we have assumed identical 
diffusion constants $D$ for both species).  Of course, once the shifts 
${\hat a} = 1 + {\bar a}$ and ${\hat b} = 1 + {\bar b}$ are performed, 
this mass term disappears (as it should), and on the classical level 
the mean-field rate equation (\ref{annmfrB}) is recovered.  It is, 
however, convenient to work with the unshifted field theory 
(\ref{cpafth}), as it again has the simple property that there are no 
(UV-divergent) Feynman diagrams that could lead to a renormalization of 
either of the propagators.  This immediately implies that the ``mass'' 
$\sigma$ is {\em not} renormalized.  Thus, as opposed to the case of 
coupled DP, there is no new non-trivial scaling field here.  If we 
further assume that there is no non-trivial contribution from the 
scaling function (i.e. unlike the case of coupled DP), then we may thus 
just insert the true density decay results (\ref{annden}) into the 
mean-field relation (\ref{slden}), in order to obtain the asymptotic 
decay for the $B$ particles.  The result is that the decay exponents 
are halved on the second hierarchy level.  The obvious generalization 
to level $i$ is therefore
\begin{equation}
\label{cpaden}
n_i(t) \sim \left\{ \begin{array}{ll} t^{-d/2^i} &{\rm for \ } d < 2 \
, \\ \left( t^{-1} \ln t \right)^{1/2^{i-1}} &{\rm for \ } d = d_c = 2
\ , \\ t^{-1/2^{i-1}} &{\rm for \ } d > 2 \ , \end{array} \right.
\end{equation}
which, with the above assumption, is an {\em exact} result at
sufficiently long times.

\begin{figure}
  \epsfxsize=55mm \centerline{\epsffile{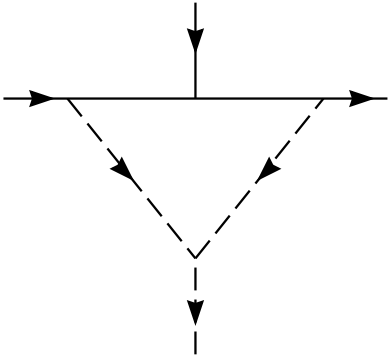}} \vspace{2mm}
\caption{
\label{FigIRAN}
Coupled annihilation: An IR-divergent diagram contributing to the $a a
{\bar a}{\bar b}$ vertex at one-loop order.}
\end{figure}

However, in a simulation with {\em finite} particle numbers, again one
would asymptotically expect a crossover to the decoupled scaling
regime, namely when there emerge large regions depleted of the $A$
species. Correspondingly, perhaps, one should notice that the coupled
annihilation problem is plagued by IR-divergent diagrams which are
very similar in nature to the coupled DP case. For example, to
one-loop order, the newly generated $a a {\bar a} {\bar b}$ vertex for
the massless shifted fields includes a diagram with two massless $a$
and two massless $b$ propagators, as depicted in
Fig.~\ref{FigIRAN}. This loop integral is infrared-singular whenever
$d\leq 6$.  One possible interpretation of these additional,
apparently non-renormalizable IR singularities, could be that they
reflect an eventual {\em non-universal} crossover to the decoupled
regime.  Neither, though, can we exclude the possibility that
ultimately a very different scaling regime ensues, which would have
to be addressed by means of an effective resummation of the expansion
with respect to the {\em relevant} coupling $\sigma$.

\begin{figure}
\epsfxsize=85mm \centerline{\epsffile{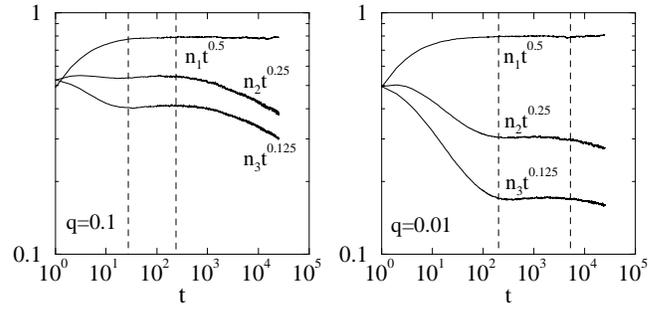}} \vspace{2mm}
\caption{
\label{FigAnnh}
The coupled annihilation process in $1+1$ dimensions: the graphs show
the densities $n_i(t) t^{1/2^i}$ of the first three levels as a
function of time for different values of the coupling strength $q$.
The scaling regime is marked by the two dashed lines (see text). Time
is measured in Monte Carlo steps.}
\end{figure}

Numerical~simulations of coupled annihilation pro\-cesses can be
performed on the same lattice as in Fig.~\ref{FigBondDP}.  The
stochastic rules have to be chosen in such a way that a particle at
site $i$ jumps to one of the neighboring sites with equal probability.
When two particles meet at the same place, they annihilate
instantaneously. Surprisingly, in the limit of infinite coupling
(instantaneous transfer of activity to the next subsystem) the
resulting curves in a log-log plot are not straight and do not
reproduce the result of Eq.~(\ref{cpaden}).  A more detailed analysis
reveals that the magnitude of these deviations depends strongly on the
coupling strength between the subsystems. To this end we replace the
instantaneous transfer of activity by a probabilistic rule, i.e.,
active particles create particles in the next subsystem at the same
location with probability $q$. Clearly, $q$ plays the role of the
parameter $\sigma$ in the field theory. By varying $q$ we observe
that the prediction of Eq.~(\ref{cpaden}) is only valid in a {\em
limited} scaling regime.  As $q$ decreases, the size of the scaling
regime grows, as illustrated in Fig.~\ref{FigAnnh}. On the other hand,
the initial crossover into the scaling regime also grows with~$q$.
Similar simulations in $3+1$ dimensions for maximal $q$ suggest that
these deviations still persist above the critical dimension although
they are much less pronounced in that case.  This supports the
conjecture that the breakdown of the scaling regime is caused by
IR-singular diagrams related to additional powers of $\sigma$, which 
would even invalidate the simple mean-field approach.

Concluding this section, we note that novel critical behavior does not
necessarily arise in the unidirectional coupling of stochastic
processes.  A counterexample is given by the following variant of
coupled annihilation, where we replace the reaction (\ref{anncp}) with
\begin{equation}
\label{anncpl}
A + A \to B + B \quad {\rm with \ rate \ } \lambda_{AB} \ .
\end{equation}
The ensuing coupled diffusion-limited reaction processes are a special
case of the more general system where the back reaction $B + B \to A +
A$ is present as well.  The full system was studied by field-theoretic
means in Ref.~\cite{howard}.  Through an analysis of the coupled
Bethe-Salpeter equations for the four non-linear vertices, it was
shown in \cite{howard} that the $A$ and $B$ reactions asymptotically
decouple, and each particle species decays according to
Eq.~(\ref{annden}).  The physical reason for this is of course that
two particles are required to meet in order for the coupling reaction
(\ref{anncpl}) to take place.  Thus, this reaction competes with the
annihilation process itself, and, in addition, as the daughter $B$
particles appear on the same sites, they have a high probability to
annihilate again immediately.  This is somewhat related to the 
robustness of the DP universality class for {\em quadratically} 
coupled DP processes \cite{mulcdp}.

\section{\bf Summary and Discussion}

The simulations presented in the last few sections show good agreement
with the predictions of the underlying field theory for a certain
range of the parameter $\tau$ for coupled DP.  Similarly, good
agreement is also found for a range of times $t$ for coupled
annihilation (or coupled DP at criticality).  Nevertheless, deep into
the critical region the simulations show a drift in the critical
scaling exponents of the second and higher hierarchy levels, perhaps
towards their decoupled values. It is not clear, however, if this
drift will go all the way towards attaining the decoupled values of
these exponents.  Furthermore, the drift is more pronounced in the
coupled annihilation model where, by decreasing the strength of the
inter-species coupling, one can extend the range of the intermediate
power-law behavior and delay the onset of the drift.  

\ From the field-theoretical point of view we believe the drift might be
due to the increasing effect of the IR-problematic diagrams which were
identified both for the coupled DP problem as well as for the coupled
annihilation problem.  These diagrams contain higher powers of the 
{\em relevant} inter-species coupling and thus are suppressed for 
small values of this coupling.  On the other hand, they become more 
dominant for larger values of the transmutation rate, which, being a 
relevant operator, increases as one goes deeper into the critical 
region.  This might perhaps render the asymptotic field theory, for 
large inter-species coupling, non-renormalizable.  Note that the
simulations for coupled annihilation show that the drift in the value
of the exponents for the second and higher hierarchy levels persists
even for $d=3$ which is above the upper critical dimension $d_c=2$ for
the first hierarchy level.  This shows that even mean-field theory may
not be valid at $d=3$, consistent with the fact that there exist
IR-singular diagrams diverging for any $d \leq 6$. Technically, a
resummation of the power expansion with respect to $\mu$ or $\sigma$ 
would be desirable; unfortunately, a more satisfactory approach to 
this problem is not yet known.

We propose the following interpretation of this scenario: eventually
we expect a {\em non-universal} crossover into decoupled behavior.
This is because in a real system, due to the discreteness of the
number of particles, which is always an integer, there are likely to
be large regions with no $A$ particles at all, and there the $B$
particles will behave as if they are decoupled from the $A$ species.
This is also true for higher hierarchy levels.  Thus we have an
interesting case in which the field theory predicts correctly the
scaling in an {\em intermediate universal} critical regime, but
eventually breaks down deep in the asymptotic limit. There is of
course the possibility that one might construct a meaningful field
theory for the asymptotic regime which will describe a crossover to a
different critical behavior, distinct from both the intermediate
regime and the decoupled behavior, but this seems a difficult task and
perhaps even an unlikely scenario at this time.

\begin{acknowledgments}
  We benefited from discussions with J.L. Cardy, A. Duncan,
  M.R. Evans, E. Frey, Y. Frishman, H.K. Janssen, M. Moshe, D. Mukamel 
  and A. Schwimmer.  We are indebted to an anonymous referee for very
  helpful comments and suggestions.  Y.Y.G. acknowledges support
  from the US Department of Energy (DOE), grant No. DE-G02-98ER45686. 
  He also thanks the Weizmann institute, where he started working on 
  this problem, and in particular E. Domany and D. Mukamel for their 
  kind hospitality.  U.C.T. acknowledges support from the Deutsche 
  Forschungsgemeinschaft (DFG) through a habilitation fellowship, 
  DFG-Gz. Ta 177 / 2-1,2.
\end{acknowledgments}


$^*$ Present address: Physics Department, Virginia Polytechnic
Institute and State University, Blacksburg, VA 24061--0435, U.S.A.

\end{document}